\documentclass{aastex}

\newcommand\arcpt{${{\lower3pt\hbox{$^{\prime\prime}$}}\atop{\raise4pt\hbox{.}}}$}

\slugcomment{to appear in the {\it Astronomical Journal}}

\shorttitle{New High Proper Motion Stars}
\shortauthors{Subasavage et al.}

\begin{document}

\title{The Solar Neighborhood XII: \\ Discovery of New High Proper
Motion Stars \\ with 1.0$\arcsec$/yr $>$ $\mu$ $\ge$ 0.4$\arcsec$/yr
between Declinations $-$90$^\circ$ and $-$47$^\circ$ }

\author{John P. Subasavage, Todd J. Henry}

\affil{Georgia State University, Atlanta, GA 30303--3083}

\author{Nigel C. Hambly}

\affil{Institute for Astronomy, University of Edinburgh \\ Royal
Observatory, Blackford Hill, Edinburgh, EH9~3HJ, Scotland, UK}

\author{Misty A. Brown and Wei-Chun Jao}

\affil{Georgia State University, Atlanta, GA 30303--3083}

\email{subasavage@chara.gsu.edu}


\begin{abstract}

We report the discovery of 141 new high proper motion systems
(1.0$\arcsec$/yr $>$ $\mu$ $\ge$ 0.4$\arcsec$/yr) in the southern sky
($\delta$ $=$ $-$90$^\circ$ to $-$47$^\circ$) brighter than UKST plate
$R_{59F}$ $=$ 16.5 via our SuperCOSMOS-RECONS (SCR) search.  When
combined with the nine systems having $\mu$ $\ge$ 1.0$\arcsec$/yr
and/or late spectral type from the initial phases of this effort~
\citep{2004AJ....128..437H,hen04}, we find that 73 of the 150 total
systems are moving faster than 0.5$\arcsec$/yr, and are therefore new
members of the classic ``LHS'' (Luyten Half Second) sample.  These
constitute a 21\% increase in the sample of stars with $\mu$ $\ge$
0.5$\arcsec$/yr in the declination region searched, thereby comprising
an important addition to this long-neglected region of the sky.

Distance estimates are provided for the entire sample, based upon a
combination of photographic plate magnitudes and 2MASS photometry,
using the relations presented in \citet{2004AJ....128..437H} for the
presumed main sequence stars.  Three systems are anticipated to be
within 10 pc, and an additional 15 are within 25 pc.  Eight of these
18 nearby systems have proper motions falling between 0.4$\arcsec$/yr
and 0.6$\arcsec$/yr, hinting at a large population of nearby stars
with fast, but not extremely high, proper motions that have not been
thoroughly investigated.

\end{abstract}


\section{Introduction}

In an effort to identify the Sun's nearest neighbors, the Research
Consortium on Nearby Stars (RECONS) is utilizing the SuperCOSMOS Sky
Survey (SSS) to reveal previously unknown high proper motion (HPM)
stars in the southern hemisphere.  Here we report the discovery of 141
new HPM systems with 1.0$\arcsec$/yr $>$ $\mu$ $\ge$ 0.4$\arcsec$/yr
found between declinations $-$90$^\circ$ and $-$47$^\circ$ that are
brighter than $R_{59F}$ of 16.5.  The stars are mainly white dwarfs,
subdwarfs, and red dwarfs that are underrepresented in the solar
neighborhood population because of their intrinsic faintness
\citep{1997AJ....114..388H}.

Recent HPM surveys, listed in Table~\ref{pm-surveys}, have uncovered
numerous new stars that complement the traditional surveys of
\citet{1971lpms.book.....G,1978LowOB...8...89G} and
\citet{1979lccs.book.....L}.  Luyten summarized his effort in the
famous Luyten Half Second Catalogue \citep[][hereafter
LHS]{1979lccs.book.....L}, which has become the gold standard for more
recent proper motion surveys.  While most of these surveys have
targeted limited pieces of the sky, all have revealed important new
HPM objects (the work of \citet{2003A&A...397..575P} is currently
difficult to assess because our initial checking of many of the
objects in the list of 6206 detections indicates that many are
previously known and several are not real HPM sources).  Noteworthy is
the large SUPERBLINK survey employed by
\citet{2002AJ....124.1190L,2003AJ....126..921L} in the northern
hemisphere, which included 49\% of the entire sky, and consequently
ranks as the largest contributor of new LHS stars since the pioneering
days of Luyten and Giclas.

One goal of our RECONS group is to complete a comprehensive proper
motion survey of the southern sky that reaches to a magnitude limit
similar to that of the L{\' e}pine work (mag $\sim$20 at $R$).  To
reveal new high proper motion objects, we mine the SuperCOSMOS
database developed and maintained at the Royal Observatory in
Edinburgh, Scotland.  Two previous papers in {\it The Solar
Neighborhood} series, \citet{2004AJ....128..437H} and \citet{hen04},
report initial results of this effort, which we refer to as the SCR
(SuperCOSMOS-RECONS) survey.  In this paper we present comprehensive
results of the SCR survey for the portion of sky centered on the south
celestial pole and reaching northward to $\delta$ $=$ $-$47$^\circ$.

\section {Search Methodology}

The search techniques utilized here are identical to those in
\citet{2004AJ....128..437H}, where a full discussion can be found.
The SCR search utilizes all astrometric and photometric information
from the four photographic plates available ($B_J$, $ESO-$$R$,
$R_{59F}$, and $I_{IVN}$) in the portion of the sky searched.
Parameter limits for the current search are 10.0$\arcsec$/yr $\ge$
$\mu$ $\ge$ 0.4$\arcsec$/yr and brighter than $R_{59F}$ $=$ 16.5.

The current search includes 13.4\% of the entire sky reaching from the
south celestial pole northward to $\delta$ $=$ $-$47$^\circ$.  As
shown in Figure~\ref{pltfields}, a few fields have not been searched
because of a limited spread in epochs for available plates or crowding
near the Magellanic Clouds or Galactic plane.  These missed regions
include only 2.3\% of the entire sky, so the current SCR search covers
83\% of the sky south of $\delta$ $=$ $-$47$^\circ$.

In the search region, a total of 1424 candidate objects were detected
with the adopted parameters.  A three-step sifting process was then
used to vet the candidates for true and false detections, including
checks of magnitudes, colors, and image ellipticities: (1) the two $R$
magnitudes were checked for consistency, and (2) the colors were
examined to determine whether they matched that of a real object,
i.e., both $B-R$ and $B-I$ positive, or both negative.  If the
candidate passed these initial two checks, it was selected for visual
inspection.  In cases where a candidate failed the first two tests,
(3) the ellipticity quality flag was also checked.  Experience
revealed that if two or more image ellipticities were larger than 0.2,
the object was spurious.  Detections that failed all three tests were
classified as false without visual inspection.  As a final check, all
of the 99 candidates found between $\delta$ $=$ $-$90$^\circ$ and
$-$80$^\circ$ were inspected visually (regardless of the checks), and
all fell into the appropriate true or false detection bins.

For the true detections, coordinates were cross-checked with the
SIMBAD database and the NLTT \citep{1995yCat.1098....0L} catalog.  If
the coordinates agreed to within a few arcminutes and the magnitudes
and proper motion were consistent, the detection was considered
previously known.  In a few cases, the coordinates and proper motions
agreed well, but the magnitudes did not.  Three of these near matches
turned out to be new common proper motion companions to previously
known proper motion objects.

The final count of real, distinct, new systems with 1.0$\arcsec$/yr
$>$ $\mu$ $\ge$ 0.4$\arcsec$/yr and brighter than $R_{59F}$ $=$ 16.5
is 150, including five systems from \citet{2004AJ....128..437H} and
four additional systems from \citet{hen04}.  For completeness, all 150
objects are listed in Table~\ref{scr-tbl} and finder charts are given
at the end of this paper in Figure~\ref{finders}.  We continue using
our naming convention, ``SCR'' for objects discovered during the
survey.

It is worth noting that the extension of the cutoff from
1.0$\arcsec$/yr in \citet{2004AJ....128..437H} down to 0.4$\arcsec$/yr
in this paper has resulted in an increased hit rate for objects ---
only 0.7\% of objects detected with 10.0$\arcsec$/yr $\ge$ $\mu$ $\ge$
1.0$\arcsec$/yr are real, while 87\% of objects detected with
1.0$\arcsec$/yr $>$ $\mu$ $\ge$ 0.4$\arcsec$/yr are real.  These
fractions include both new and known objects.  The higher hit rate is
due to the fact that higher proper motion searches are susceptible to
spurious contamination because reliable source association between
different epochs is more difficult for fast-moving sources.  The
greater the search area for the counterpart, the higher the likelihood
that an incorrect match will be made (especially if one plate has a
defect, a fairly common occurance).

\section {Comparison to Previous Proper Motion Surveys}

The classic work of Luyten still remains the most fruitful proper
motion survey to date.  This is demonstrated in Figure~\ref{luyten},
in which the objects listed in the LHS Catalogue with $\mu$ $\ge$
0.5$\arcsec$/yr are plotted.  Many HPM objects were also provided by
\citet{1971lpms.book.....G,1978LowOB...8...89G}, although none were
south of $\delta$ $=$ $-$47$^\circ$.  What is immediately obvious is
that objects south of $\delta$ $=$ $-$30$^\circ$ are undersampled
relative to similar northern declinations.  A direct comparison of
counts in the four quartiles of sky illustrate this bias clearly: 1004
objects are found between $\delta$ $=$ $+$90$^\circ$ and
$+$30$^\circ$, 988 between $+$30$^\circ$ and 00$^\circ$, 944 between
00$^\circ$ and $-$30$^\circ$, but only 666 between $-$30$^\circ$ and
$-$90$^\circ$.  It is therefore not surprising that six of the seven
recent proper motion surveys have concentrated on the southern
hemisphere.  The one notable exception is the work of L{\' e}pine and
collaborators \citeyearpar{2002AJ....124.1190L,2003AJ....126..921L}
utilizing SUPERBLINK, which has been very successful at filling in
gaps in the northern hemisphere (primarily along the Galactic plane).
Table~\ref{pm-surveys} lists the number of new discoveries by each of
these surveys.  Of the southern hemisphere surveys, many only cover
small portions of the sky.  Although we avoid the Galactic plane and
Magellanic Clouds, the SCR survey has the most uniform sky coverage in
the southern hemisphere.  Figure~\ref{surveys} shows the distribution
of HPM objects discovered by five recent proper motion surveys
(distinguished by various symbols; \citet{2003A&A...397..575P} not
shown because of the difficulties mentioned previously).

A primary goal of the SCR effort is to further complete the LHS
Catalogue for stars listed by Luyten to have $\mu$ $\ge$
0.5$\arcsec$/yr.  Our extension of the cutoff to $\mu$ $\ge$
0.4$\arcsec$/yr in this survey is to ensure that no known LHS stars
were missed due to proper motion measurement errors for objects very
near the 0.5$\arcsec$/yr limit.  Only two objects for which our
measured proper motion was below 0.5$\arcsec$/yr are LHS stars --- LHS
3694 with $\mu$ $=$ 0.493$\arcsec$/yr, and LHS 3803 with $\mu$ $=$
0.444$\arcsec$/yr.

An assessment of the completeness of the SCR search indicates that for
$R_{59F}$ $<$ 16.5, we recover 216 of 287 (75\%) of LHS stars in the
portion of sky searched.  For stars brighter than $R_{59F}$ $=$ 10.0,
we recover only 29 of 71 (41\%) of LHS stars because the search is
insensitive to bright objects that are saturated in the photographic
emulsions.  In what we consider the ''sweet spot'' of the SCR search,
10.0 $<$ $R_{59F}$ $<$ 16.5, we recover 87\% of the LHS stars.  This
relatively high recovery rate is virtually identical for stars moving
faster than 1.0$\arcsec$/yr (30 of 35) or 0.4--1.0$\arcsec$/yr (155 of
178), indicating that there is no particular bias in whether fast or
slow moving objects are recovered more easily.

In addition to LHS recoveries, the SCR search recovered numerous
objects from other recent proper motion surveys.  During the
compilation of values listed in Table~\ref{pm-surveys}, we noticed
sources with similar sets of coordinates appearing in more than one
survey.  In total, seven stars have been found to be duplicates, even
though two surveys claimed the discovery.  Table~\ref{pm-surveys}
reflects counts for each object only once, assigned to the original
discovery survey.  The seven overlap stars include: ER 2
\citep{1987RMxAA..14..381R} is listed as WT 392
\citep{1991A&AS...91..129W}; LHS 1140, LHS 1147, LHS 1152, and LHS
1160 \citep{1979lccs.book.....L} are listed as WT 1138, WT 1147, WT
1161, and WT 1170 \citep{1996A&AS..115..481W} respectively; LHS 3983
\citep{1979lccs.book.....L} is listed as WT 1007
\citep{1994A&AS..105..179W}; WT 1141 \citep{1996A&AS..115..481W} is
listed as WD0045-061 \citep{2001Sci...292..698O}.

\section {Data Mining}
\subsection {SuperCOSMOS --- Astrometry and Plate Photometry}

Coordinates, proper motions, and plate magnitudes have been extracted
from SuperCOSMOS, and are listed in Table~\ref{scr-tbl}.  Coordinates
are for epoch and equinox J2000.  Errors in the coordinates are
typically $\pm$ 0.3$\arcsec$, and errors in the proper motions are
given.  Errors in position angle are usually $\pm$ 0.1$^\circ$.
Photometric magnitudes are given for three sets of plates --- $B_J$,
$R_{59F}$, and $I_{IVN}$.  Extractions of a region around each SCR
object have been made to check for problems in the data retrieved via
the automated source extractions.  In a few cases, one or more plates
were not available or sources were merged, thereby preventing the
determination of reliable magnitudes.

\subsection {2MASS --- Infrared Photometry}

Infrared photometry is used to extend the color baseline, which allows
more accurate photometric distance estimates for red dwarfs and
permits a fairly reliable separation of the white and red dwarfs.  The
infrared $JHK_s$ photometry has been extracted from 2MASS via Aladin.
Each SCR object has been identified by eye to ensure that no extracted
magnitudes are in error.  In nearly every case, the errors are smaller
than 0.03 mag.  Exceptions include objects with $J$ $>$ 15, $H$ $>$
14.5, and $K_s$ $>$ 14, where the errors are 0.05 mag or greater.  In
several cases where $H$ $>$ 16 or $K_s$ $>$ 15, the error is null, and
the value is therefore unreliable.

\section {Analysis}
\subsection {Color-Magnitude Diagram}

The color magnitude diagram illustrated in Figure~\ref{colmag} clearly
shows that the new SCR objects are generally fainter and redder than
those found by previous studies.  Sources south of $\delta$ $=$
$-$47$^\circ$ with $\mu$ $\ge$ ~0.5$\arcsec$/yr from the previous
studies are collectively labeled as ``known'' in Figure~\ref{colmag}
(and in Figure~\ref{redpromo} as well).  Large open symbols for the
SCR stars indicate new LHS members, while smaller symbols are for
additional SCR stars with 0.5$\arcsec$/yr $>$ $\mu$ $\ge$
0.4$\arcsec$/yr.

Other than the white dwarfs, there are very few known sources fainter
than $R_{59F}$ $\sim$ 14, whereas most of the new SCR sources are
fainter.  This is caused primarily by the lack of red plates available
to Luyten.  Nonetheless, there are a half dozen new SCR objects
brighter than $R_{59F}$ $=$ 13, the brightest of which has $R_{59F}$
$=$ 11.7.  That such bright HPM objects remain unknown indicates that
this portion of the sky, which has comprehensive coverage only from
the Bruce Proper Motion Survey carried out by Luyten (with a blue
photographic limit of $\sim$ 15.5), had not yet been thoroughly
searched until this SCR effort.

Several of the objects are quite red, including the remarkable object
SCR 1845-6357 \citep[$V-K_s$ $=$ 8.89, M8.5V,][]{hen04} for which a
trigonometric parallax of $\pi$ $=$ 282 $\pm$ 23 mas has been
determined from photographic plates \citep{dea05}.  This object is
represented in Figure~\ref{colmag} as the single point to the far
right.  We note that there is no obvious decrease in the number of
objects at the faint limit of $R_{59F}$ $=$ 16.5 adopted for the
current search, hinting that there is likely to be a large population
of fainter objects yet to be discovered in the SuperCOSMOS data.

\subsection {Reduced Proper Motion Diagram}

The reduced proper motion (RPM) diagram shown in Figure~\ref{redpromo}
is a powerful diagnostic for assigning rough luminosity classes for
stars using the observables, proper motion and apparent magnitude.  It
is similar to an H-R diagram except that absolute magnitude is
replaced by reduced proper motion, where the proper motion is used in
lieu of a trigonometric parallax measurement to determine $H_R$, as
follows:

\begin{displaymath}
H_R = R_{59F} + 5 + 5\log\mu.
\end{displaymath}

The assumption here is that proper motion is directly related to
distance --- an assumption certainly not always valid because high
velocity populations, such as subdwarfs with large tangential
velocities, can masquerade as very nearby main sequence stars.
Nonetheless, utilization of the RPM diagram allows the identification
(albeit roughly) of subdwarfs, which are found at fainter $H_R$ values
for a given color or, alternately, at bluer colors for a given $H_R$.
In addition, white dwarfs are clearly differentiated from the main
sequence stars and subdwarfs, providing a reliable means for
identifying new white dwarfs.

From Figure~\ref{redpromo} it is apparent that most of the new SCR
stars are main sequence red dwarfs, as expected, while there are a few
dozen new subdwarf candidates.  The dotted line represents a somewhat
arbitrary boundary between the subdwarfs and white dwarfs, of which
five new candidates have been found in the SCR search.  As in
Figure~\ref{colmag}, the single point to the far right is SCR
1845-6357.

\subsection {Red Dwarfs --- Photometric Distances}

Once all available data has been collected, we can search for targets
that are potentially nearby, specifically within the volumes defined
by the RECONS sample (horizon at 10 pc) and the CNS \citep[Catalog of
Nearby Stars,][]{1991adc..rept.....G} or NStars (Nearby Stars) samples
(horizons at 25 pc).

Distances for the SCR objects in Table~\ref{scr-tbl} have been
estimated using a combination of plate magnitudes from SuperCOSMOS and
infrared photometry from 2MASS, following the methodology of
\citet{2004AJ....128..437H}.  Briefly, the six magnitudes provide 15
color--M$_K$ combinations, 11 of which can be used to estimate
individual distances ($JHK_s$--only colors are not used because of
limited color discrimination, and $(B_J - R_{59F})$ is not reliable).
These relations have been developed using a large set of stars within
10 pc that have high quality parallaxes, plate magnitudes
(specifically extracted from the SuperCOSMOS database for this
purpose), and 2MASS infrared photometry.  The relations assume that
the objects are single, main sequence, dwarfs of types $\sim$K0V to
M9V.  To estimate the reliability of this technique, the same 10 pc
sample that generated the relations was run through giving a mean
difference between the photometric and trigonometric distances of
26\%.

Results indicate that three of the 150 systems are within 10 pc (each
has been discussed previously in \citealt{2004AJ....128..437H} and/or
\citealt{hen04}), SCR 1845-6357 (3.5 pc), SCR 0630-7643AB (6.9 pc),
and SCR 1138-7721 (8.8 pc).  An additional 15 systems are predicted to
be within 25 pc.  

We have also run distance estimates for the 266 known objects
recovered and find that 120 have distance estimates placing them
within the 25 pc horizon.  Of these, 34 objects have trigonometric
parallaxes in the Yale Parallax Catalogue \citep[hereafter
YPC]{1995gcts.book.....V} and/or from the {\it Hipparcos} mission
\citep{1997yCat.1239....0E} that confirm they are within 25 pc, while
24 objects have trigonometric parallaxes placing them beyond 25 pc.
As with any volume-limited survey, most of the objects are found near
the distance limit, so it is not surprising that a substantial sample
of objects slip beyond the adopted horizon.  These statistics strongly
suggest the necessity of determining a trigonometric parallax prior to
inclusion of the NStars 25 pc sample based solely on photometric
distance estimates.  The remaining 62 stars have no trigonometric
parallax.  Of these, 21 are currently being observed as part of our
Cerro Tololo InterAmerican Observatory Parallax Investigation (CTIOPI)
Program including three with distance estimates nearer than 10 pc: LHS
271 (GJ 1128, 8.0 pc), LHS 263 (GJ 1123, 8.2 pc), and LHS 532 (GJ
1277, 9.1 pc).

A small subsample of 16 SCR stars have distance estimates in excess of
200 pc (the five white dwarf candidates have been removed from this
count).  K and M type subdwarfs tend to have larger distance estimates
than their true distances because the relations assume luminosities of
main sequence stars, whereas subdwarfs are intrinsically fainter
(hence closer for a given brightness).  These 16 stars are therefore
the best subdwarf candidates.  Worthy of note is the most extreme star
in this subsample, SCR 2249-6324, which has a distance estimate of
nearly 400 pc.  The colors of this object indicate that it is only
slightly red, though red enough to nearly eliminate the possibility
that it is a cool white dwarf.  Its position in Figure~\ref{redpromo}
($H_R$ $=$ 19.56, $(R_{59F} - J)$ $=$ 1.58) lies in the subdwarf
region just above the broken line.  Follow-up spectroscopy will
confirm its luminosity class.

\subsection {White Dwarfs}

Five SCR discoveries lie in the white dwarf region of
Figure~\ref{redpromo} --- SCR 0252-7522, SCR 0311-6215, SCR 0821-6703,
SCR 2012-5956, and SCR 2016-7945.  We have obtained spectroscopy on
each and confirmed that all are white dwarfs.  The spectra will be
presented in a future publication.

Of the 150 systems in the complete list, only two have no distance
estimate due to colors too blue to be addressed by the relations ---
SCR1257-5554B and the white dwarf SCR 0311-6215.  Cooler white dwarfs
having colors that are within the ranges covered by the relations tend
to have very large distance estimates because the relations assume
they are intrinsically bright F or G dwarfs.  The distance estimates
in Table~\ref{scr-tbl} for the four new white dwarfs with colors
covered by the relations range from 267 pc to 734 pc.  This provides a
second diagnostic for identifying white dwarfs that is useful but not
as reliable as the reduced proper motion diagram.

SCR 1257-5554B is a source too faint to be picked up in this SCR
search, but was noticed on frames that were blinked to confirm its
primary.  Infrared data is not available because this object exceeds
the faint limit of 2MASS.  The $R$ magnitude quoted in
Table~\ref{scr-tbl} is $ESO-R$ due to blending in the $R_{59F}$ plate.
We suspect it is a hot white dwarf due to its plate colors and because
its companion is a modestly bright M star estimated to be at 39 pc.
The B component is not plotted in Figure~\ref{redpromo} due to the
lack of the $(R_{59F} - J)$ color.

\citet{2001Sci...292..698O} has derived a single color linear fit to
obtain distance estimates for white dwarfs using plate magnitudes.  We
utilize this relation and adopt the error quoted therein of 20\% to
give the following distances: SCR 0252-7522 $=$ 29.8 $\pm$ 6.0 pc; SCR
0311-6215 $=$ 60.7 $\pm$ 12.2 pc; SCR 0821-6703 $=$ 10.9 $\pm$ 2.2 pc;
SCR 2012-5956 $=$ 18.0 $\pm$ 3.6 pc (consistent with \citet{hen04}
using CCD photometry and the relation of \citet{2004ApJ...601.1075S}
to obtain a distance of 17.4 $\pm$ 3.5 pc); SCR 2016-7945 $=$ 29.3
$\pm$ 5.9 pc.  Note that should the distance for SCR 0821-7522 hold
true, it would become an addition to the 13 pc sample (thought to be
largely complete) from which \citet{2002ApJ...571..512H} determine the
white dwarf local density.  CTIOPI parallax observations are currently
underway for verification.

\subsection {Comments on Individual Systems}

The five systems with $\mu$ $\ge$ 1.0$\arcsec$/yr have been discussed
in detail in \citet{2004AJ....128..437H} and \citet{hen04}.  Four more
systems (one double) having spectral types of M6.0V or M6.5V were also
discussed in \citet{hen04}.  Here we provide details of additional
noteworthy systems, each of which is a multiple.

SCR 0005-6103 ($\mu$ $=$ 0.504$\arcsec$/yr at position angle 084.3
degrees) is a common proper motion companion to LHS 1018 ($\mu$ $=$
0.519$\arcsec$/yr at position angle 085.7 degrees), for which there is
no trigonometric parallax available.  The distance estimates for SCR
0005-6103 and LHS 1018 are 43.2 pc and 34.0 pc, respectively.  This is
a reasonably good match considering the errors in the distance
estimation technique (26\%).

SCR 0006-6617 ($\mu$ $=$ 0.559$\arcsec$/yr at position angle 161.7
degrees) at first appears to be a very widely separated
($\sim$ 27$\arcmin$) common proper motion companion to LHS 1019 ($\mu$
$=$ 0.576$\arcsec$/yr at position angle 158.9 degrees).  However, the
estimated distance for SCR 0006-6617 is 63.2 pc, while the
trigonometric parallax from YPC for LHS 1019 indicates a distance of
17.6 pc.  In addition, SCR 0006-6617 ($H_R$ $=$ 18.85, $(R_{59F} - J)$
$=$ 3.10) does not fall within the subdwarf region of
Figure~\ref{redpromo} rendering it unlikely that this object is a
subdwarf with an overestimated distance estimate.  We conclude that
this is a rare case of two physically unassociated objects of similar
proper motion being found in the same region of sky.

SCR 0630-7643AB was discussed in \citet{hen04}.  It is a new nearby
($\sim$ 7 pc) binary with separation 1.0$\arcsec$ and brightness
difference of $\sim$0.25 mag at $I_C$.

SCR 1257-5554AB was discussed in $\S$5.4 as a probable red dwarf/white
dwarf pair.

SCR 2155-7330 is a common proper motion companion to HIP 108158.  The
Hipparcos parallax for this object is 0.02510 $\pm$ 0.00074$\arcsec$
(distance $=$ 39.8 pc), which is reasonably consistent with the
photometric distance estimate for SCR 2155-7330 of 31.6 pc.

SCR 2250-5726AB is noticeably peanut-shaped in the SuperCOSMOS frames.
CCD frames taken at the CTIO 0.9m confirm it to be a close binary
source with separation 2.3$\arcsec$ at position angle 28$^\circ$.

SCR 2352-6124 is a common proper motion companion to LHS 4031, which
has a Hipparcos parallax of 0.02070 $\pm$0.00120$\arcsec$ (distance
$=$ 48.3 pc), which is consistent with our photometric distance
estimate for SCR 2352-6124 of 50.3 pc.

\section {Discussion}

Listed in Table~\ref{diststats} is a summary of the number of SCR
systems with distance estimates within each of the two horizons (10 pc
and 25 pc) and beyond.  The five confirmed white dwarfs have been
removed; however, there remain several likely subdwarfs with
overestimated distances that are not accounted for in the statistics
--- a few may be within 25 pc.  Perhaps the most surprising result of
this survey is the discovery that the slowest proper motion bin of
width 0.2$\arcsec$/yr --- stars moving between 0.4$\arcsec$/yr and
0.6$\arcsec$/yr --- contains the largest number (nine) of new
candidates for systems within 25 pc.  One of these systems, SCR
0630-7643AB, is a binary with separation $\sim$1$\arcsec$ that is
probably only $\sim$7 pc distant \citep{hen04}, yet has a relatively
low proper motion of only 0.483$\arcsec$/yr.  Of course, the largest
number of new proper motions stars is found in the slowest bin, yet
the presence of so many nearby candidates hints that large numbers of
nearby stars may lie undetected at even smaller proper motion values.
It is quite possible that stars within a few parsecs of the Sun have
escaped detection simply because they exhibit little proper motion.
Pushing to lower proper limits and combining efforts with full-sky
photometric surveys may yet reveal these hidden neighbors.

The SCR search detailed here for 13.4\% of the sky currently has
revealed $\sim$one-third as many new LHS systems as the SUPERBLINK
effort of \citet{2002AJ....124.1190L,2003AJ....126..921L}, which
covered 49\% of the sky.  We anticipate that as we move further
northward, the discovery rate of new objects will decrease because
Luyten, Giclas, L{\' e}pine, and several of the others listed in
Table~\ref{pm-surveys} have searched portions of these regions.
Nonetheless, viable new nearby star candidates undoubtedly remain to
be found.

Once the SCR search employing the current constraints (magnitude and
proper motion) is completed for the southern hemisphere, we intend to
carry out new searches with fainter magnitude limits (to R $\sim$ 20,
consistent with L{\' e}pine's survey) and to smaller proper motions
(perhaps to $\mu$ $\ge$ 0.18$\arcsec$/yr to match the NLTT catalog
limit).  These searches will again begin at the south Galactic pole
and progress northward.  We predict that both surveys will produce
considerable numbers of potential nearby systems and help us to
provide a more accurate census of the solar neighborhood.

\section {Acknowledgments}

Funding for the SuperCOSMOS Sky Survey is provided by the UK Particle
Physics and Astronomy Research Council.  N.C.H. would like to thank
colleagues in the Wide Field Astronomy Unit at Edinburgh for their
work in making the SSS possible; particular thanks go to Mike Read,
Sue Tritton, and Harvey MacGillivray.  The RECONS team at Georgia
State University wishes to thank NASA's Space Interferometry Mission
for its continued support of our study of nearby stars.  This work has
made use of the SIMBAD, VizieR, and Aladin databases, operated at CDS,
Strasbourg, France.  We have also used data products from the Two
Micron All Sky Survey, which is a joint project of the University of
Massachusetts and the Infrared Processing and Analysis Center, funded
by NASA and NSF.


\clearpage


\figcaption[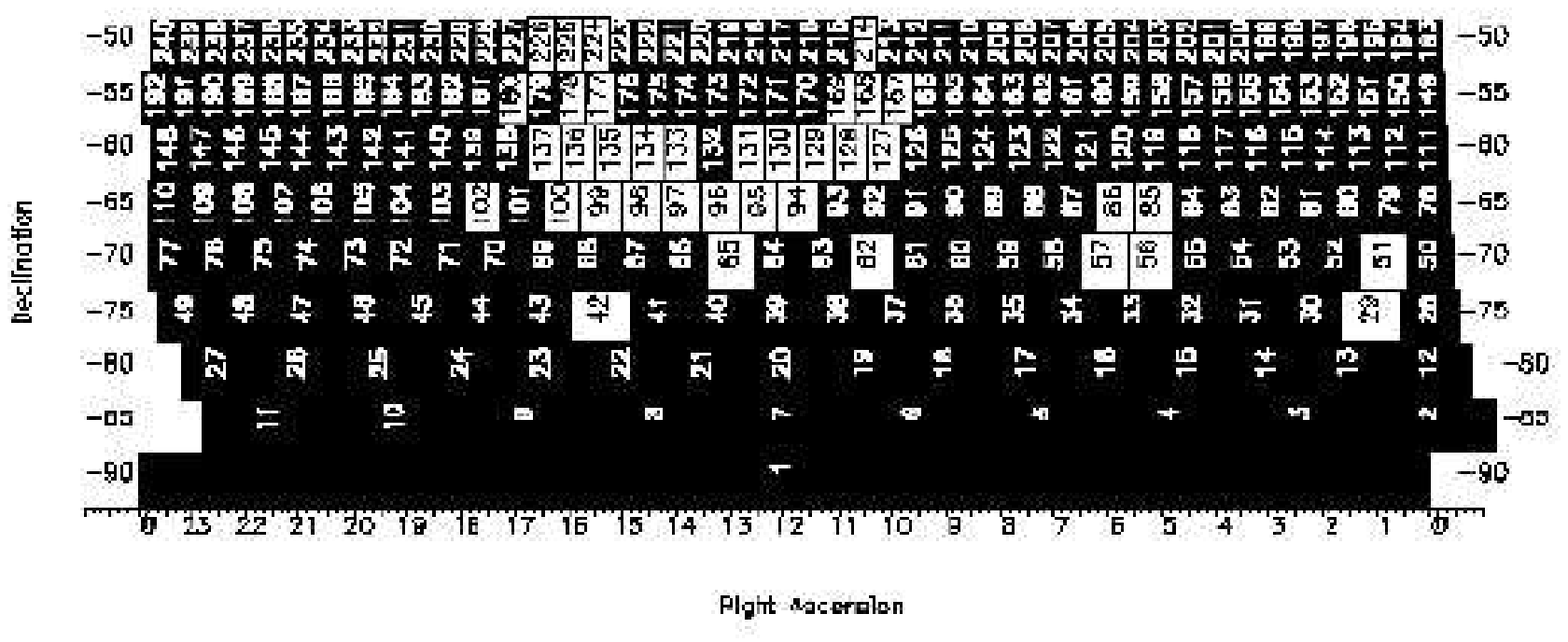]{Plate coverage of the SCR survey
reported here.  Plates colored in white were excluded from the search
for reasons mentioned in the text.
\label{pltfields}}

\figcaption[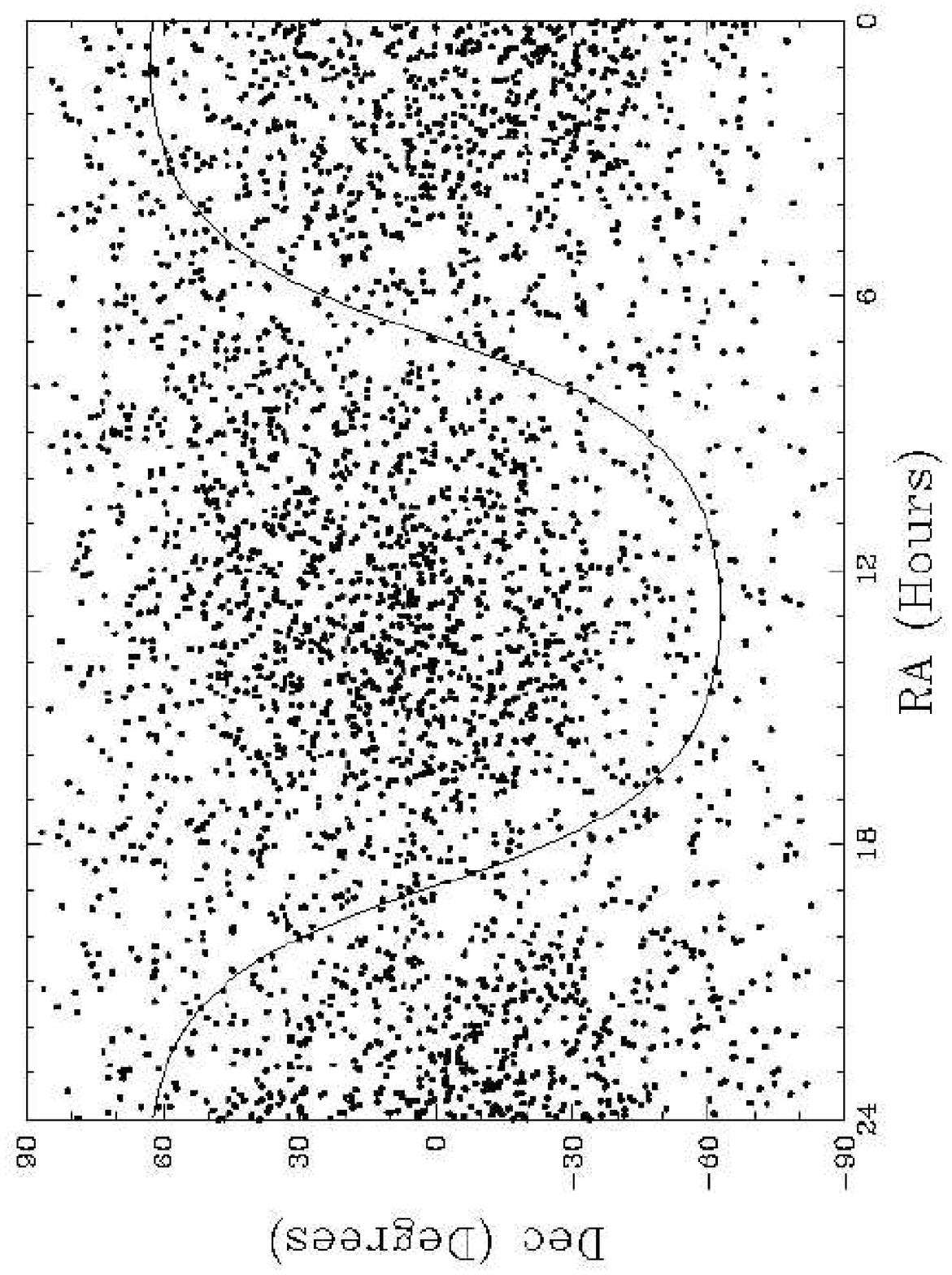]{Sky distribution of LHS objects.  Only
stars with $\mu$ $\ge$ 0.5$\arcsec$/yr are plotted.  The curve
represents the Galactic plane.
\label{luyten}}

\figcaption[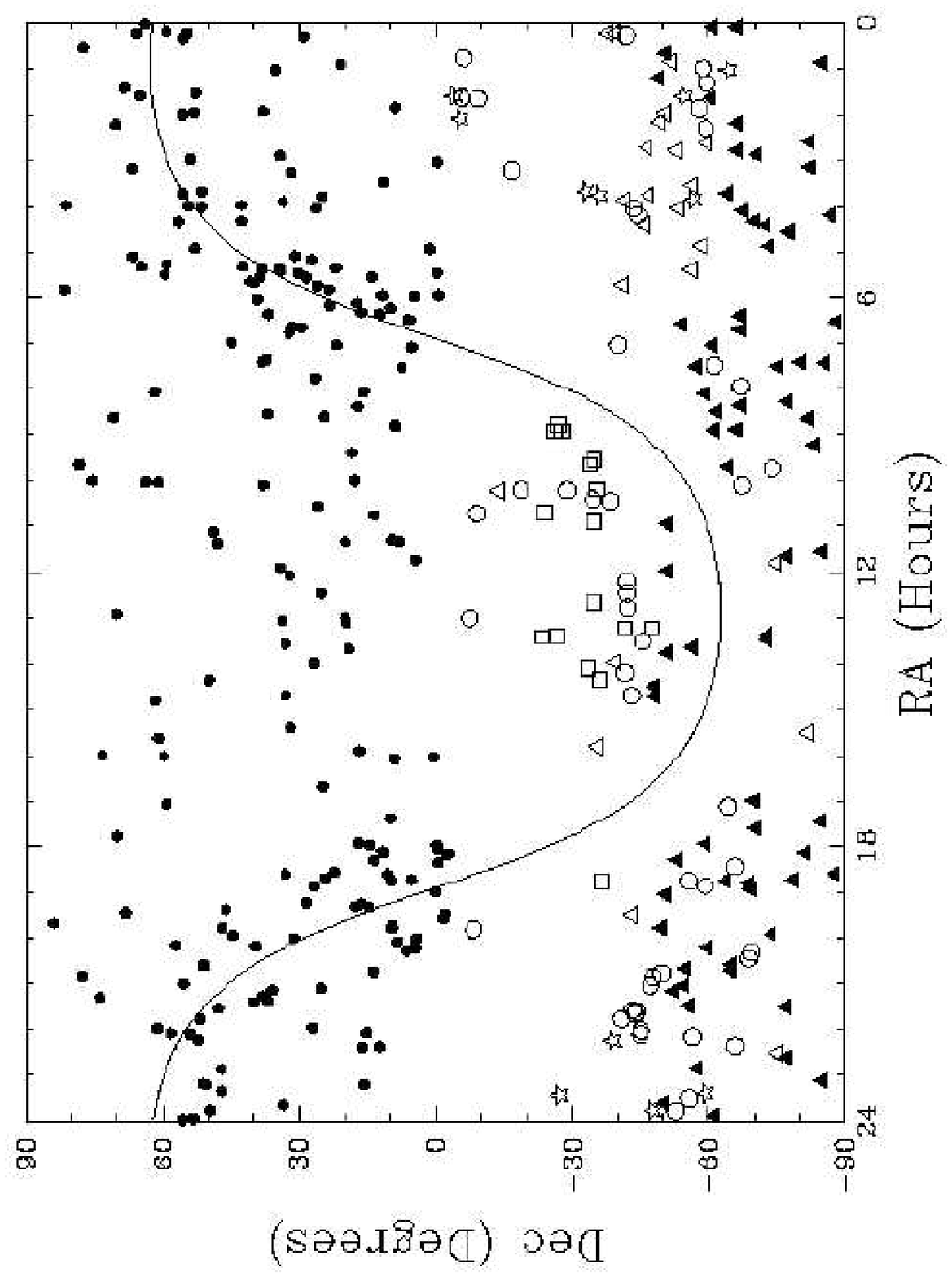]{Sky distribution of new LHS objects
from recent proper motion surveys.  Only stars with $\mu$ $\ge$
0.5$\arcsec$/yr are plotted.  Small filled circles are from the
SUPERBLINK survey.  Filled triangles are from the SCR survey.  Open
circles are from the WT survey.  Open triangles are from the Scholz
survey.  Open boxes are from the Calan-ESO survey.  Open stars are
from the Oppenheimer survey.  The curve represents the Galactic plane.
\label{surveys}}

\figcaption[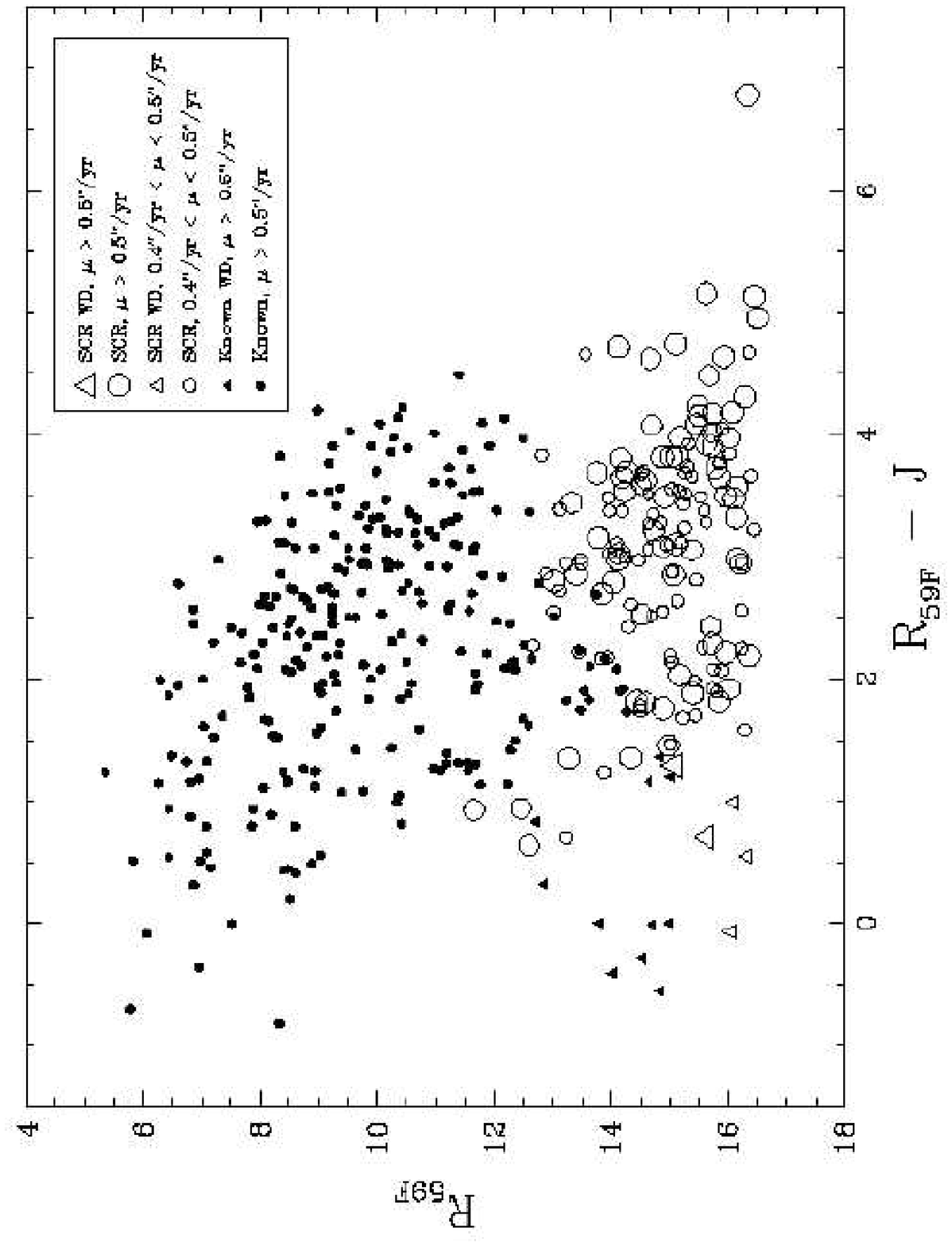]{Color-magnitude diagram for the SCR
systems with $\mu$ $\ge$ 0.4$\arcsec$/yr (size of the points splits
SCR sample into stars with $\mu$ more or less than 0.5$\arcsec$/yr)
and known systems with $\mu$ $\ge$ 0.5$\arcsec$/yr and south of
$\delta$ $=$ $-$47$^\circ$.
\label{colmag}}

\figcaption[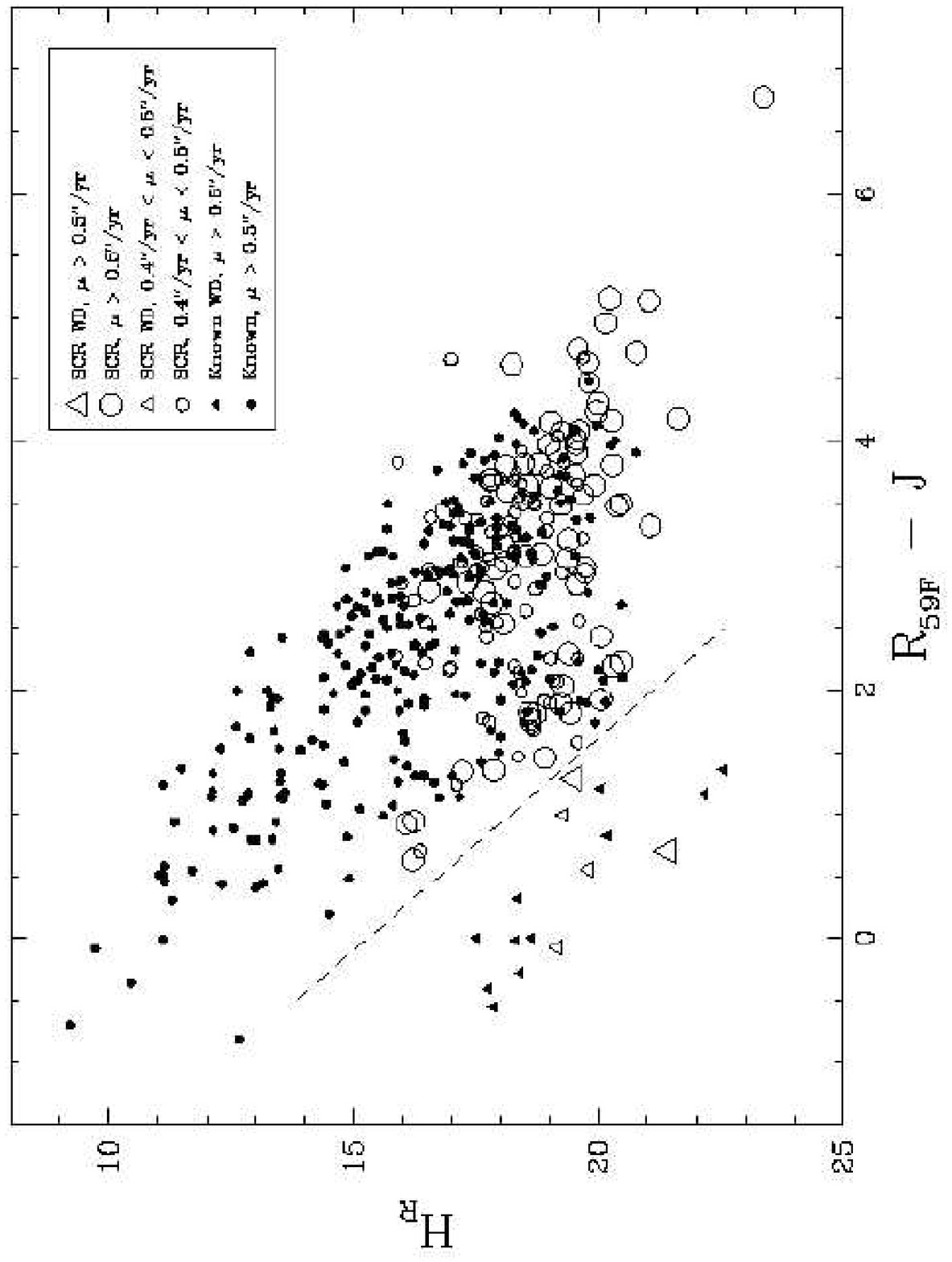]{Reduced proper motion diagram for the
SCR systems with $\mu$ $\ge$ 0.4$\arcsec$/yr (size of the points
splits SCR sample into stars with $\mu$ more or less than
0.5$\arcsec$/yr) and known systems with $\mu$ $\ge$ 0.5$\arcsec$/yr
and south of $\delta$ $=$ $-$47$^\circ$.  The dashed line serves
merely as a reference to distinguish white dwarfs from subdwarfs.
\label{redpromo}}

\figcaption[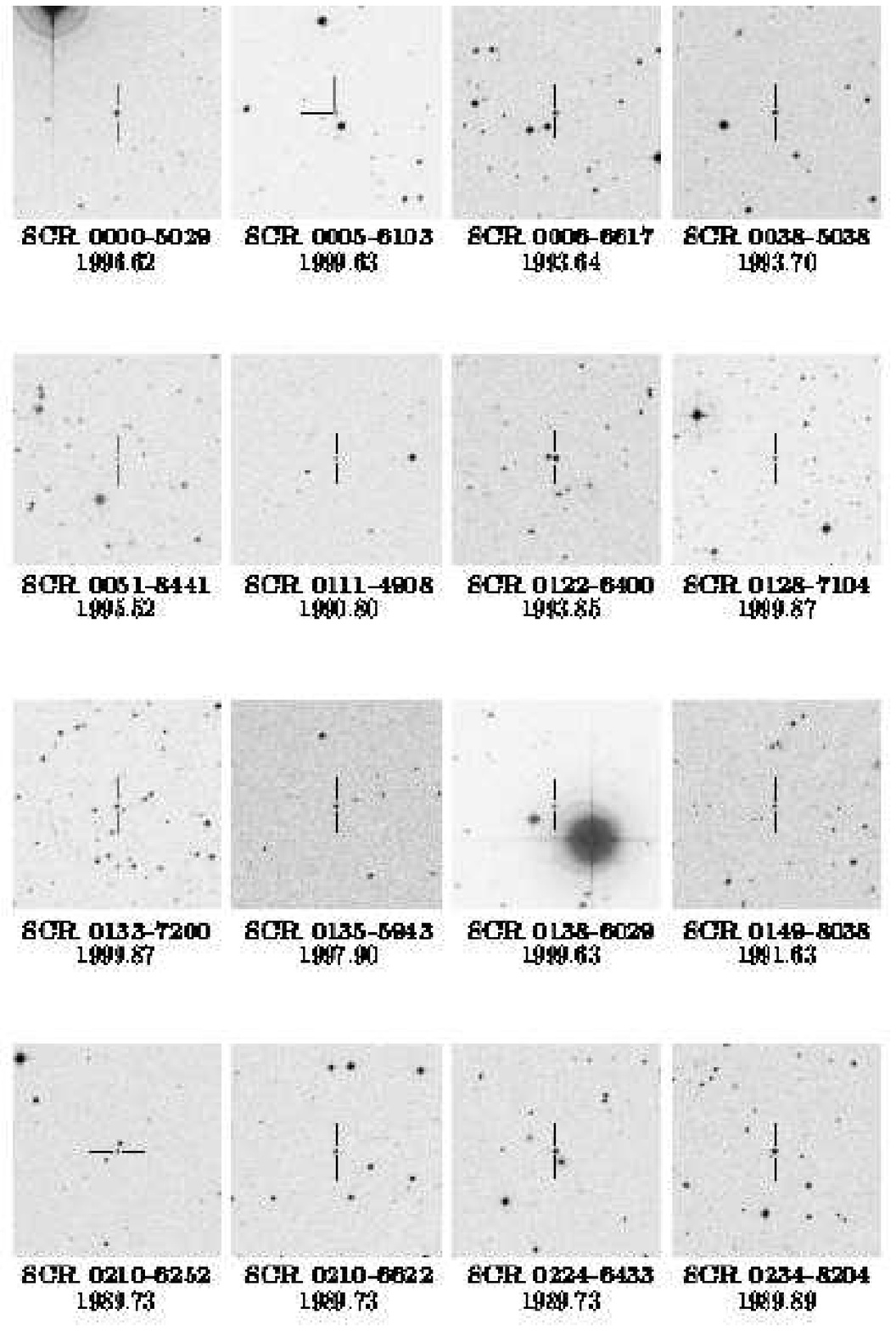]{Finder charts for the 150 SCR systems,
5$\arcmin$ on a side.  North is up, east is to the left.  The
observation epoch for each frame is given.
\label{finders}}


\begin{figure}
\plotone{subasavage.fig1.ps}
\end{figure}

\clearpage

\begin{figure}
\plotone{subasavage.fig2.ps}
\end{figure}

\clearpage

\begin{figure}
\plotone{subasavage.fig3.ps}
\end{figure}

\clearpage

\begin{figure}
\plotone{subasavage.fig4.ps}
\end{figure}

\clearpage

\begin{figure}
\plotone{subasavage.fig5.ps}
\end{figure}

\clearpage

\begin{figure}
\plotone{subasavage.fig6.ps}
\end{figure}

\clearpage

\begin{figure}
\plotone{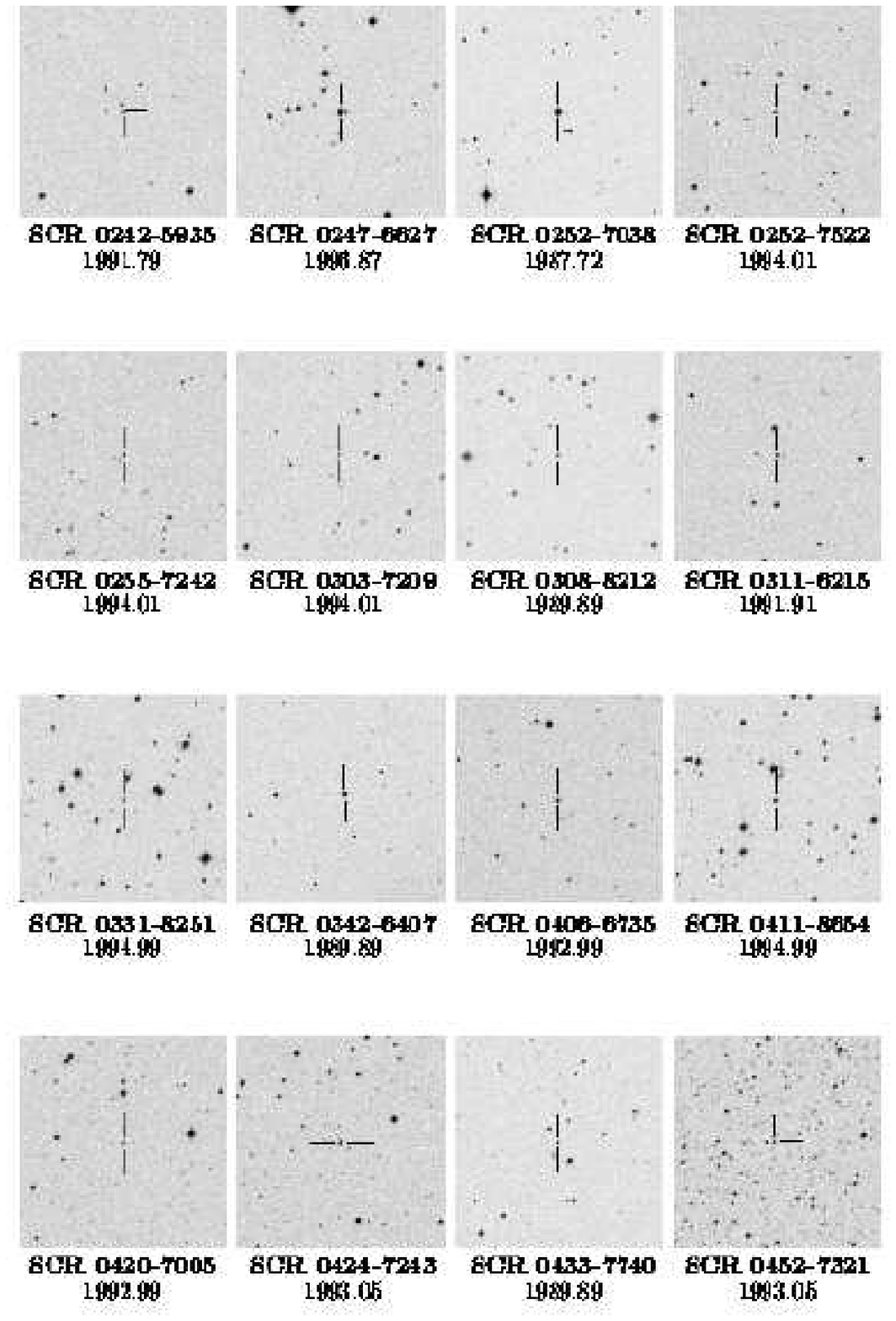}
\end{figure}

\clearpage

\begin{figure}
\plotone{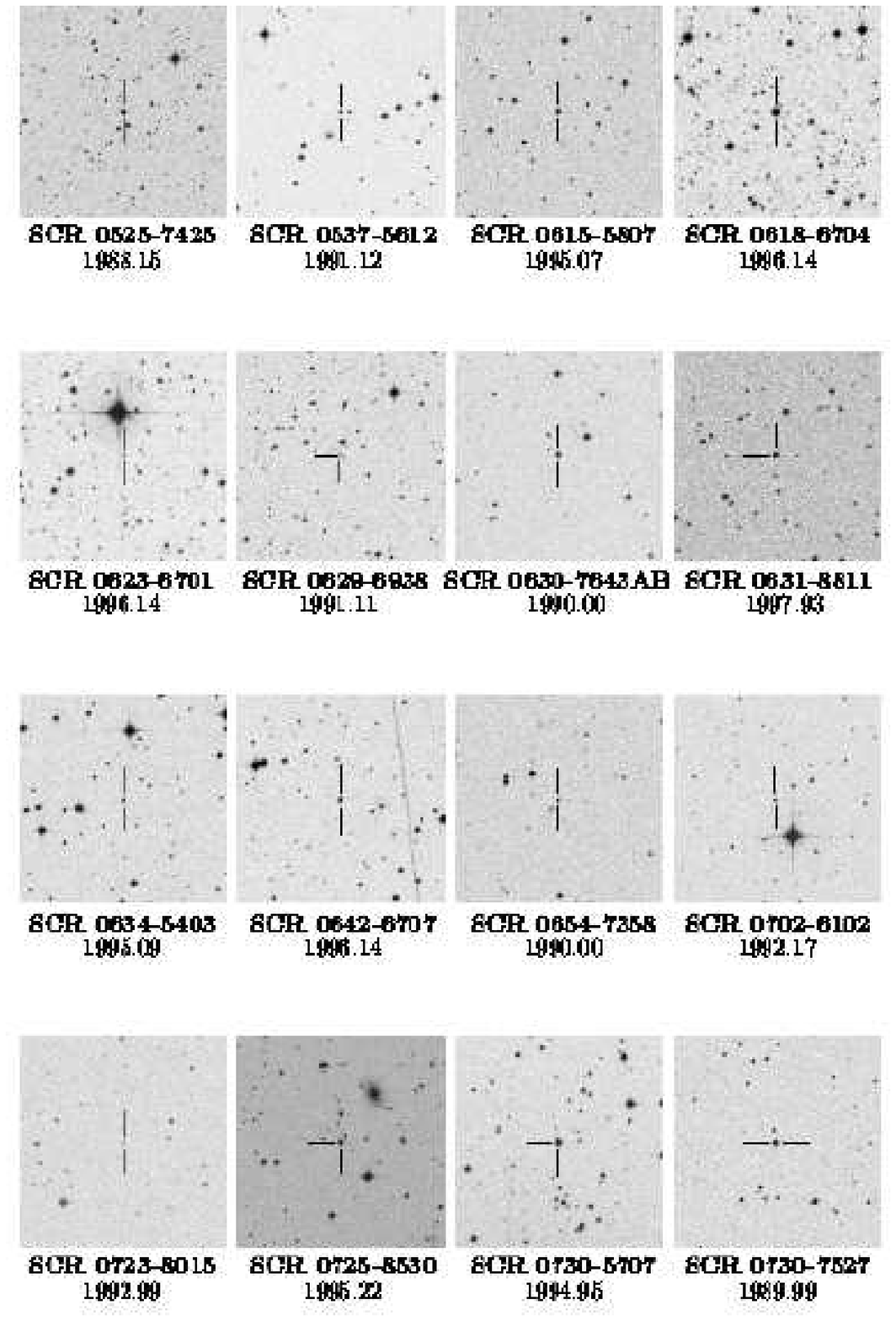}
\end{figure}

\clearpage

\begin{figure}
\plotone{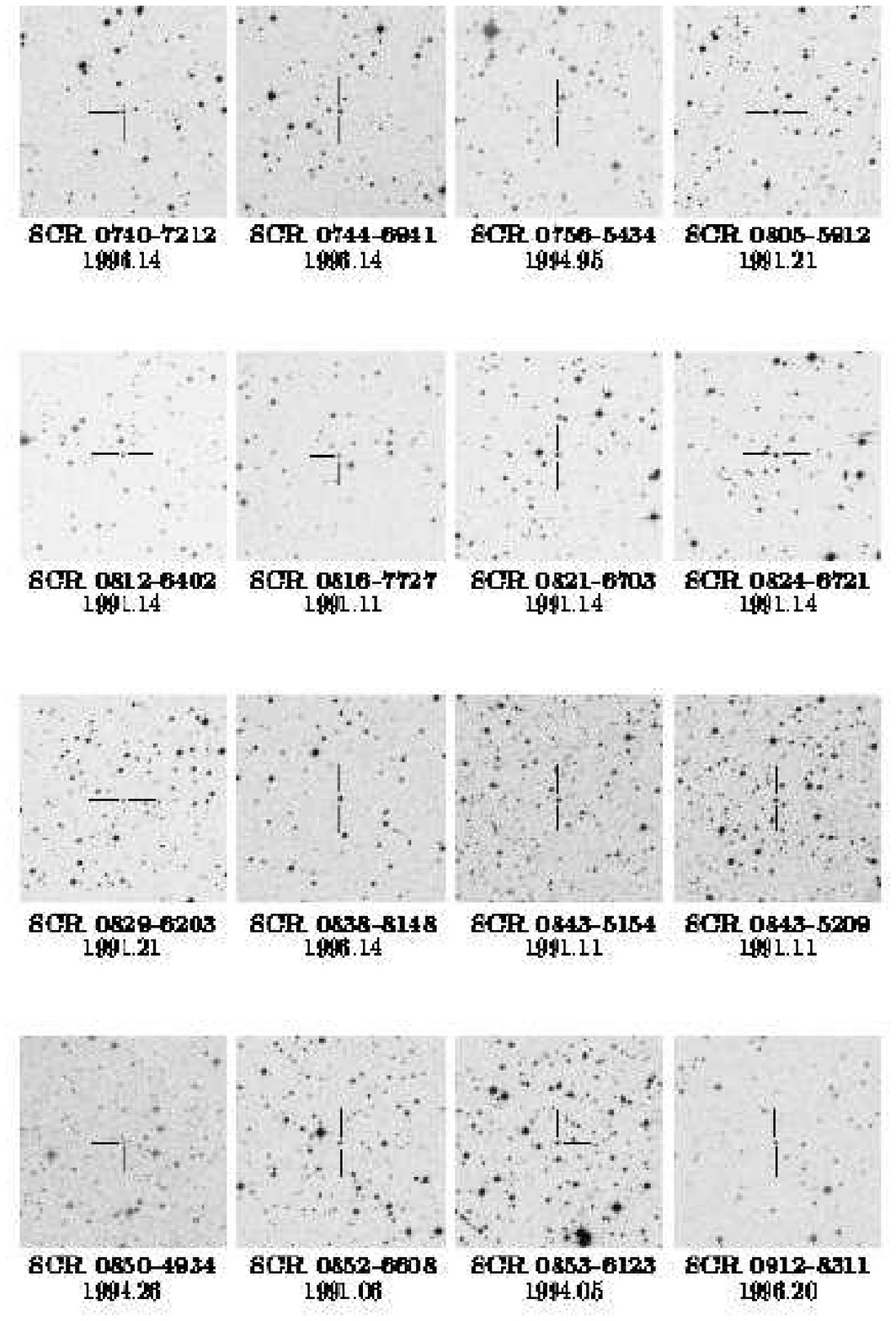}
\end{figure}

\clearpage

\begin{figure}
\plotone{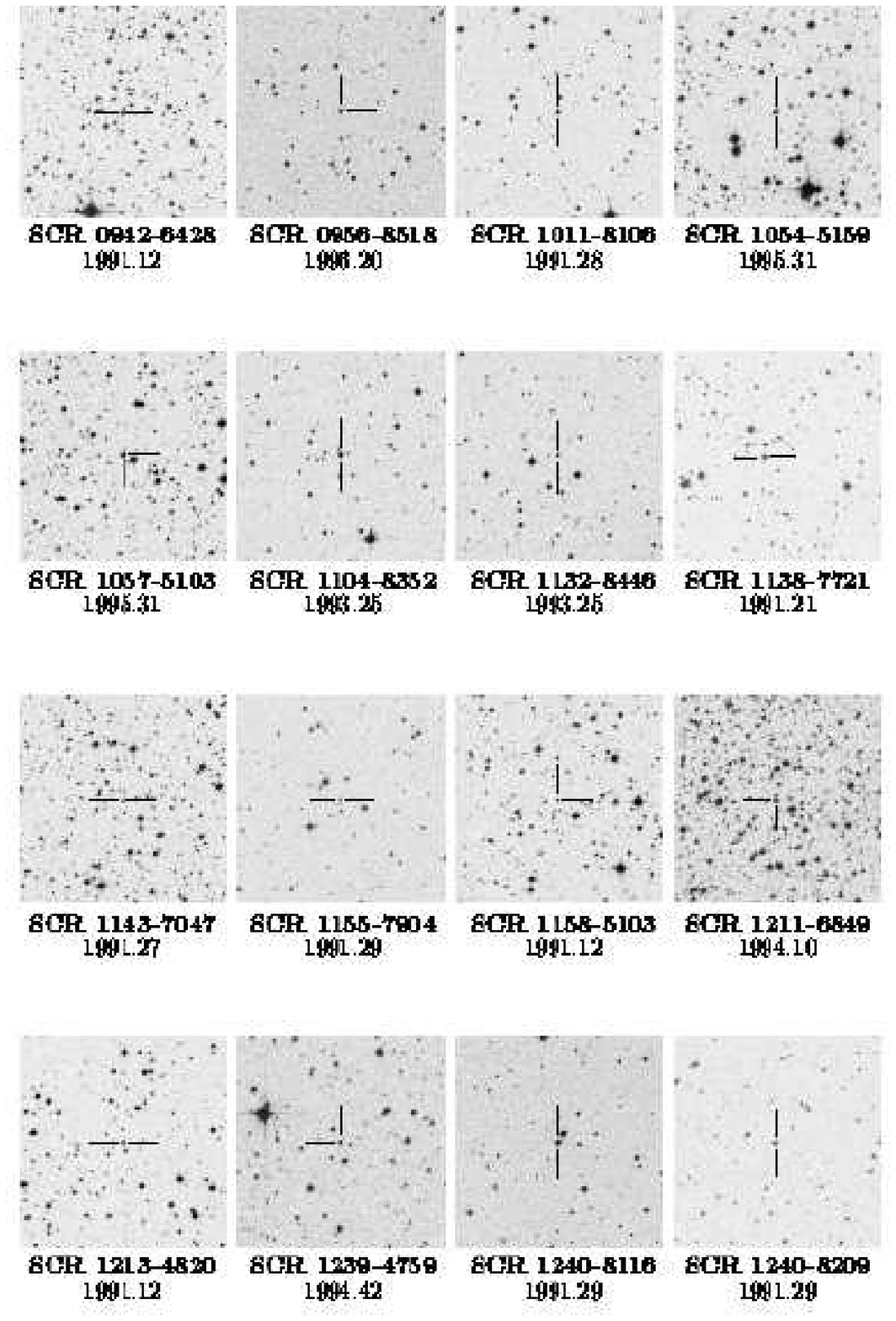}
\end{figure}

\clearpage

\begin{figure}
\plotone{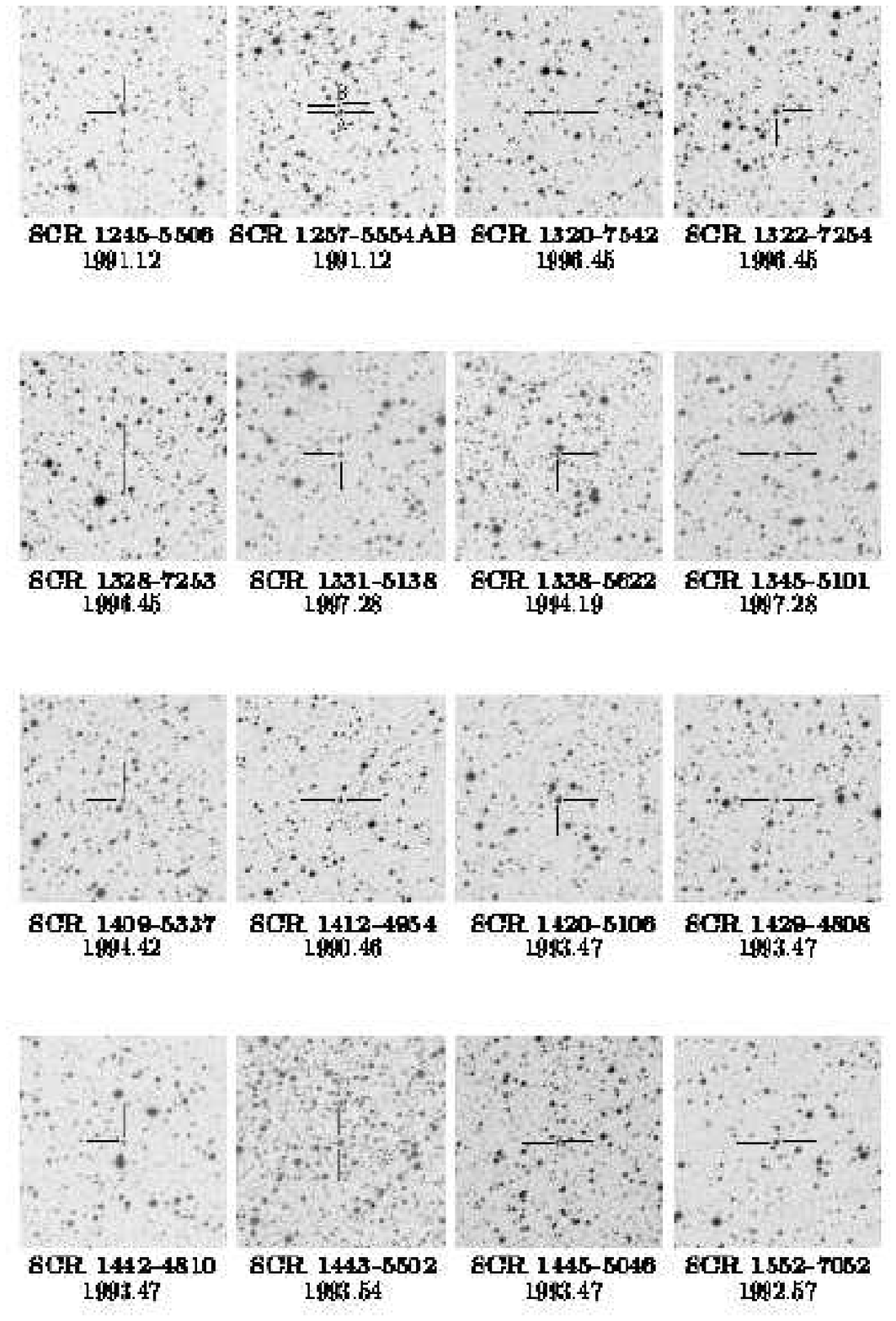}
\end{figure}

\clearpage

\begin{figure}
\plotone{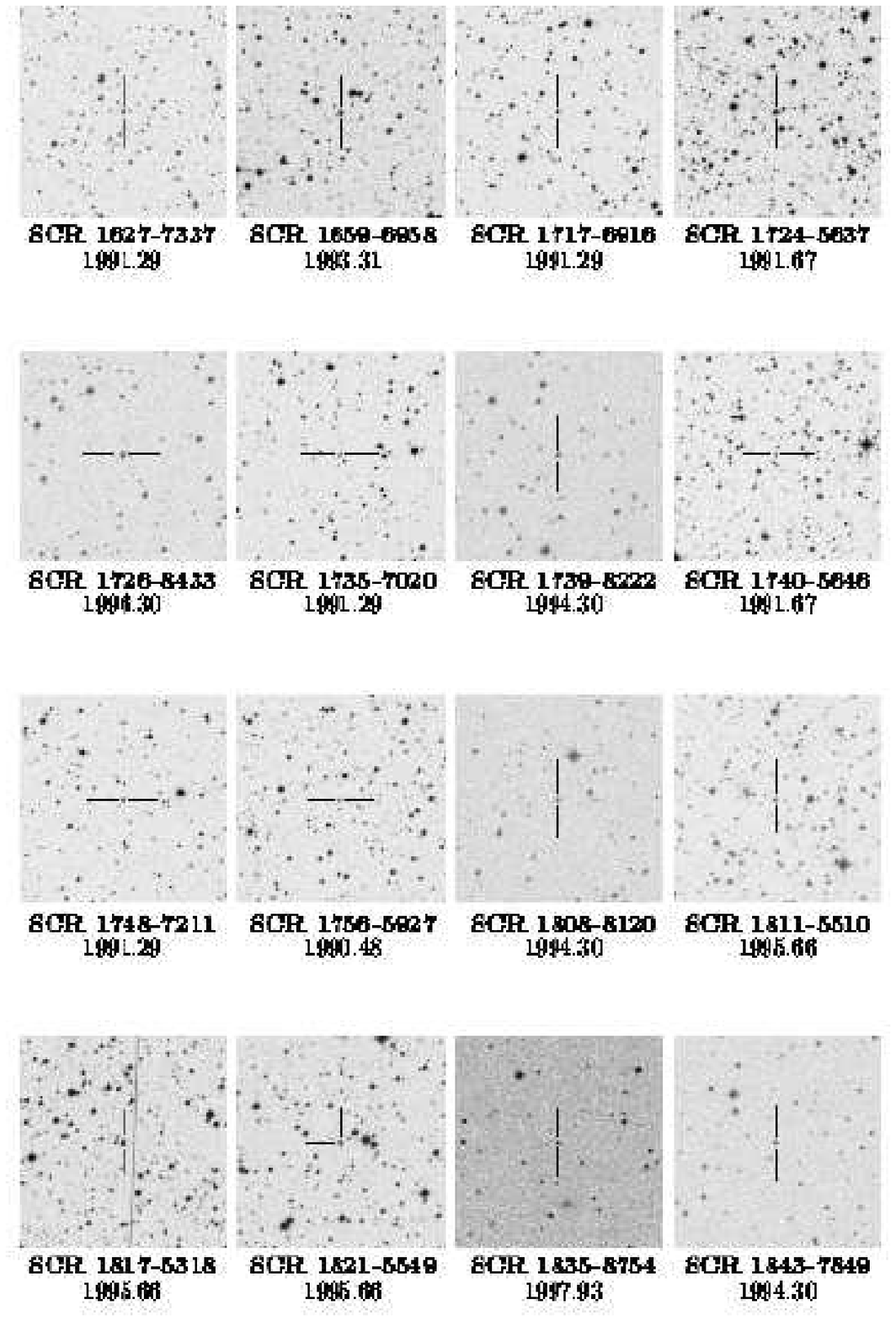}
\end{figure}

\clearpage

\begin{figure}
\plotone{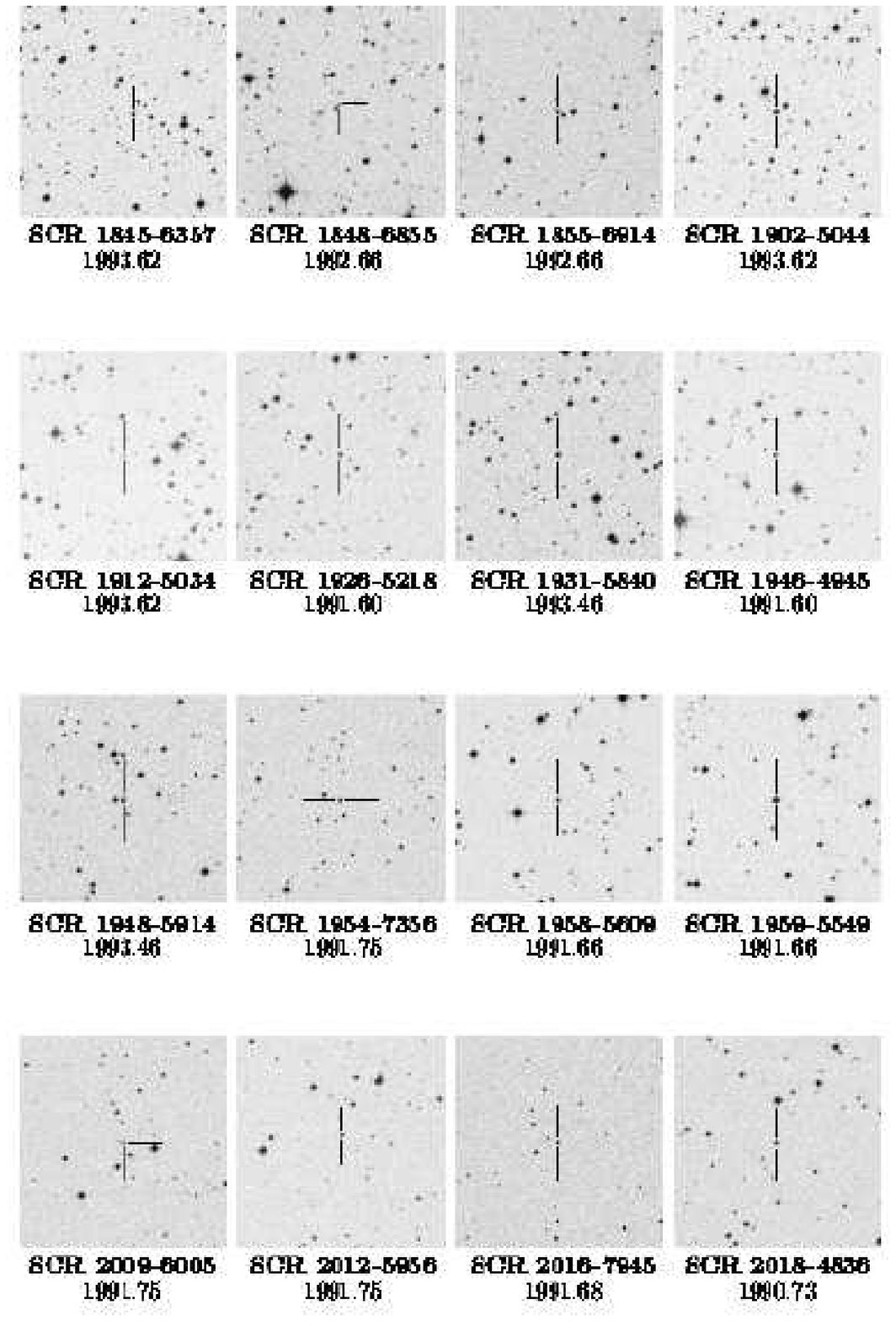}
\end{figure}

\clearpage

\begin{figure}
\plotone{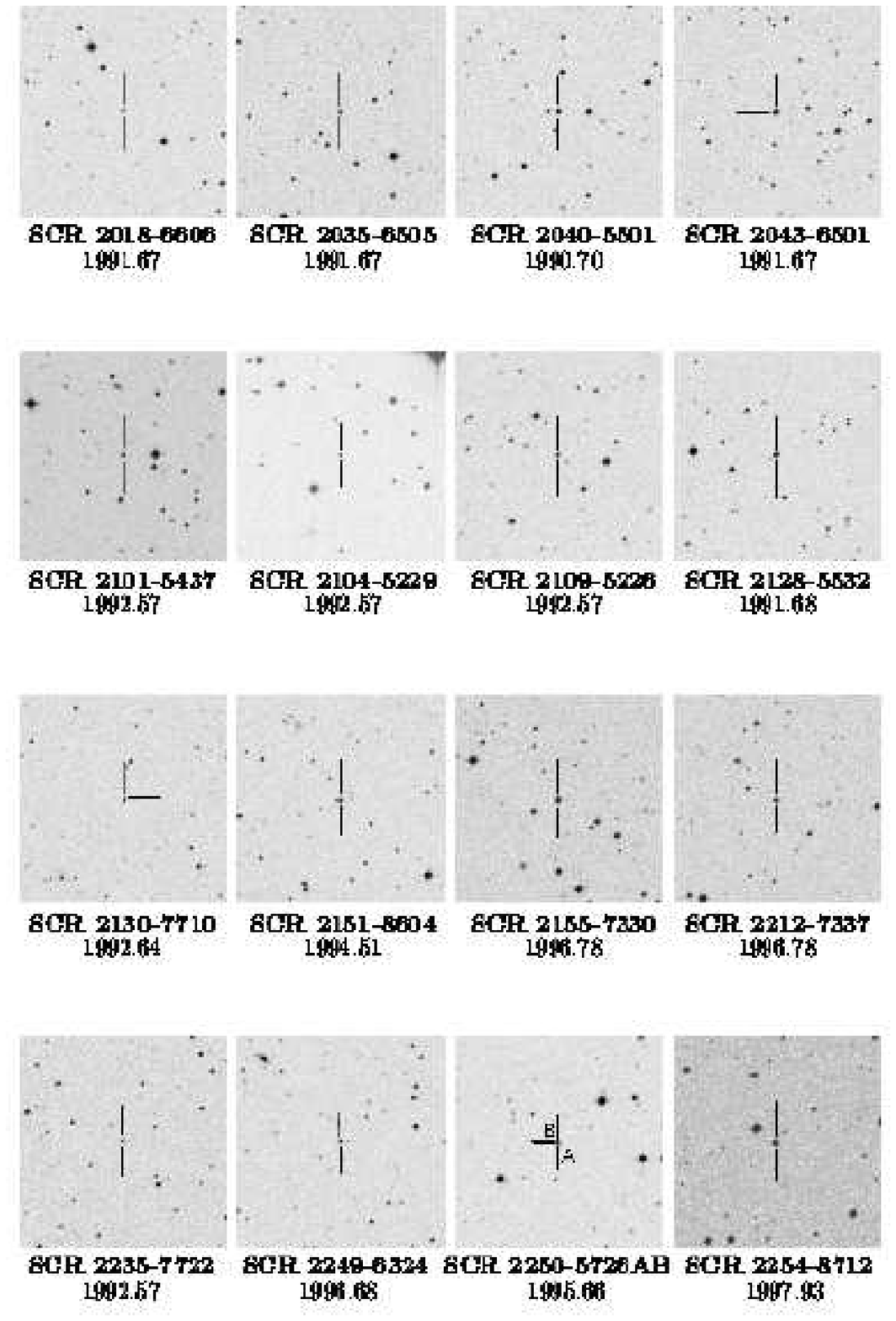}
\end{figure}

\clearpage

\begin{figure}
\plotone{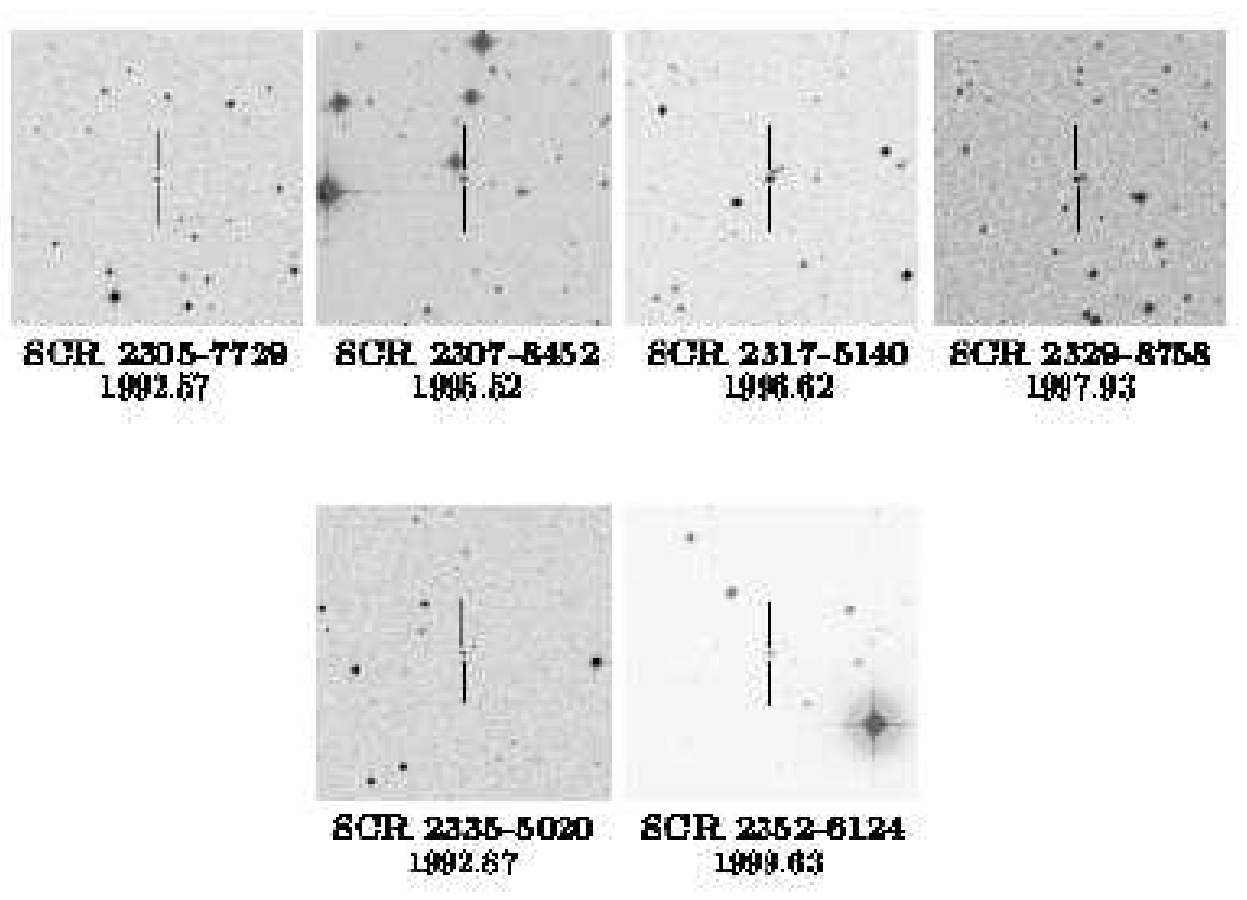}
\end{figure}

\clearpage


\begin{deluxetable}{lccc}
\tabletypesize{\footnotesize}
\tablecaption{Proper Motion Surveys and Number of New Objects Discovered. 
\label{pm-surveys}}
\tablewidth{0pt}

\tablehead{\vspace{-25pt} \\
           \colhead{Survey}&
           \colhead{$\mu$ $\ge$ 1.0$\arcsec$/yr}&
           \colhead{1.0$\arcsec$/yr $>$ $\mu$ $\ge$ 0.5$\arcsec$/yr}&
           \colhead{\# of Publications \tablenotemark{a}}}

\startdata
LHS                                &     528 &    3074 &   1   \\
SUPERBLINK                         &      18 &     180 &   2   \\
SuperCOSMOS-RECONS                 &       5 &      68 &   3   \\
WT (Wroblewski and collaborators)  &       2 &      46 &   7   \\
Scholz and collaborators           &       5 &      21 &   3   \\
Calan-ESO (Ruiz and collaborators) &       3 &      14 &   2   \\
Oppenheimer et al.                 &       3 &       8 &   1   \\
Pokorny et al.                     & unknown & unknown &   1   \\

\enddata

\tablenotetext{a}{references include \citet{1979lccs.book.....L},
\citet{2002AJ....124.1190L,2003AJ....126..921L},
\citet{2004AJ....128..437H}, \citet{hen04},
\citet{1989A&AS...78..231W,1991A&AS...91..129W,1994A&AS..105..179W,1996A&AS..115..481W,1997A&AS..122..447W},
\citet{1999A&AS..139...25W,2001A&A...367..725W},
\citet{2000A&A...353..958S,2002ApJ...565..539S,astro-ph/0406457},
\citet{1987RMxAA..14..381R,2001ApJS..133..119R},
\citet{2001Sci...292..698O}, and \citet{2003A&A...397..575P}}

\end{deluxetable}


\begin{deluxetable}{lccc}
\tabletypesize{\footnotesize}
\tablecaption{Distance Estimate Statistics for SCR Stars (Excluding White Dwarfs). 
\label{diststats}}
\tablewidth{0pt}

\tablehead{\vspace{-25pt} \\
           \colhead{Proper motion}&
           \colhead{d $\leq$ 10 pc}&
           \colhead{10 pc $<$ d $\leq$ 25 pc}&
           \colhead{d $>$ 25 pc}}

\startdata
$\mu$ $\geq$ 1.0$\arcsec$/yr                      &  2  &  0  &   2  \\
1.0$\arcsec$/yr $>$ $\mu$ $\geq$ 0.8$\arcsec$/yr  &  0  &  3  &   3  \\
0.8$\arcsec$/yr $>$ $\mu$ $\geq$ 0.6$\arcsec$/yr  &  0  &  4  &  25  \\
0.6$\arcsec$/yr $>$ $\mu$ $\geq$ 0.4$\arcsec$/yr  &  1  &  8  &  97  \\
\tableline
Total                                             &  3  & 15  & 127  \\
\enddata

\end{deluxetable}


\begin{deluxetable}{lccccrrrrrrcrc}

\rotate \tabletypesize{\scriptsize} \tablecaption{Proper motions,
photographic and infrared photometry, and distance estimates for the
SuperCOSMOS$-$RECONS sample with $\mu$ $\ge$ than 0.4$\arcsec$/yr and
south of $\delta$ $=$ $-$47$^\circ$.
\label{scr-tbl}}
\tablewidth{0pt}

\tablehead{\vspace{-25pt} \\
           \colhead{Name}&
           \colhead{RA \hskip25pt DEC}&
           \colhead{$\mu$}&
	   \colhead{$\sigma_\mu$}&
	   \colhead{$\theta$}&
           \colhead{$B_J$}&
           \colhead{$R_{59F}$}&
           \colhead{$I_{IVN}$}&
           \colhead{$J$}&
           \colhead{$H$}&
           \colhead{$K_s$}&
           \colhead{$R_{59F}$ $-$ $J$}&
           \colhead{Est Dist}&
           \colhead{Notes}\\

           \colhead{}&
           \colhead{(J2000)}&
           \colhead{($\arcsec$)}&
           \colhead{($\arcsec$)}&
	   \colhead{($^\circ$)}&
           \colhead{}&
           \colhead{}&
           \colhead{}&
           \colhead{}&
           \colhead{}&
           \colhead{}&
           \colhead{}&
           \colhead{(pc)}&
           \colhead{}}

\startdata
\vspace{-25.05pt} \\
\tableline \vspace{-15pt} \\
\multicolumn{14}{c}{SuperCOSMOS$-$RECONS sample with $\mu$ $\ge$ 0.5$\arcsec$/yr} \\
\tableline \vspace{-15pt} \\

  SCR 0005-6103  &  00 05 56.49 $-$61 03 55.2  &  0.504  &  0.009  &  084.3  &  18.19  &  16.02  &  13.37  &  12.04  &  11.43  &  11.18  &   3.98  &    43.2  & cpm with LHS 1018  \\	      	
  SCR 0006-6617  &  00 06 33.73 $-$66 17 30.8  &  0.559  &  0.011  &  161.7  &  16.94  &  15.11  &  13.18  &  12.01  &  11.36  &  11.11  &   3.10  &    63.2  &        \\			
  SCR 0038-5038  &  00 38 48.00 $-$50 38 22.3  &  0.726  &  0.010  &  115.1  &  16.77  &  14.92  &  12.76  &  11.43  &  10.94  &  10.66  &   3.49  &    44.9  &        \\			
  SCR 0051-8441  &  00 51 16.42 $-$84 41 59.0  &  0.502  &  0.007  &  077.2  &  18.22  &  16.20  &  14.57  &  13.26  &  12.71  &  12.50  &   2.94  &   126.3  &        \\			
  SCR 0111-4908  &  01 11 47.51 $-$49 08 09.0  &  0.542  &  0.008  &  213.1  &  18.93  &  16.50  &  13.01  &  11.54  &  11.00  &  10.61  &   4.96  &    23.6  &        \\			
  SCR 0138-6029  &  01 38 01.13 $-$60 29 56.0  &  0.580  &  0.009  &  083.3  &  17.46  &  15.17  &  12.44  &  11.19  &  10.66  &  10.29  &   3.98  &    28.7  &        \\			
  SCR 0210-6622  &  02 10 45.17 $-$66 22 26.6  &  0.769  &  0.020  &  056.3  &  16.86  &  14.60  &  12.15  &  10.97  &  10.43  &  10.09  &   3.63  &    30.0  &        \\			
  SCR 0234-8204  &  02 34 44.28 $-$82 04 25.3  &  0.618  &  0.017  &  333.3  &  20.29  &  14.13  &  12.81  &  11.11  &  10.51  &  10.25  &   3.03  &    28.5  &        \\			
  SCR 0247-6627  &  02 47 05.35 $-$66 27 14.3  &  0.711  &  0.016  &  053.4  &  15.94  &  13.79  &  12.15  &  10.63  &  10.10  &   9.78  &   3.16  &    31.3  &        \\			
  SCR 0252-7038  &  02 52 32.02 $-$70 38 22.3  &  0.767  &  0.013  &  201.0  &  13.56  &  11.66  &  10.86  &  10.73  &  10.23  &  10.04  &   0.92  &    69.1  &        \\			
  SCR 0308-8212  &  03 08 54.56 $-$82 12 30.6  &  0.507  &  0.009  &  027.3  &  17.80  &  15.62  &  13.09  &  11.70  &  11.15  &  10.89  &   3.92  &    38.9  &        \\			
  SCR 0342-6407  &  03 42 57.44 $-$64 07 56.4  &  1.071  &  0.023  &  141.2  &  17.17  &  15.13  &  12.34  &  11.32  &  10.89  &  10.58  &   3.81  &    39.3  & \tablenotemark{a} \\	      	
  SCR 0406-6735  &  04 06 06.79 $-$67 35 28.9  &  0.608  &  0.007  &  150.2  &  17.29  &  14.98  &  13.99  &  13.53  &  13.06  &  12.80  &   1.45  &   239.4  &        \\			
  SCR 0411-8654  &  04 11 38.07 $-$86 54 09.8  &  0.557  &  0.006  &  046.5  &  17.83  &  15.79  &  13.39  &  12.06  &  11.53  &  11.26  &   3.73  &    51.3  &        \\			
  SCR 0420-7005  &  04 20 12.54 $-$70 05 58.8  &  0.670  &  0.007  &  021.2  &  18.18  &  15.68  &  12.58  &  11.19  &  10.59  &  10.25  &   4.49  &    22.5  & \tablenotemark{b} \\	      	
  SCR 0424-7243  &  04 24 33.63 $-$72 43 04.8  &  0.563  &  0.006  &  033.5  &  16.43  &  15.03  &  12.67  &  11.21  &  10.67  &  10.40  &   3.82  &    38.0  &        \\			
  SCR 0433-7740  &  04 33 26.62 $-$77 40 09.7  &  0.514  &  0.006  &  049.6  &  17.92  &  15.86  &  14.76  &  14.05  &  13.49  &  13.36  &   1.81  &   291.4  &        \\			
  SCR 0452-7321  &  04 52 06.87 $-$73 21 56.7  &  0.554  &  0.008  &  053.6  &  17.98  &  16.29  &  13.83  &  11.98  &  11.44  &  11.12  &   4.31  &    39.1  &        \\			
  SCR 0623-6701  &  06 23 09.04 $-$67 01 18.9  &  0.514  &  0.007  &  027.9  &  \nodata&  16.15  &  13.64  &  12.58  &  12.09  &  11.81  &   3.57  &    76.3  &        \\			
  SCR 0631-8811  &  06 31 31.28 $-$88 11 36.8  &  0.516  &  0.006  &  349.9  &  16.96  &  14.67  &  11.46  &  10.04  &   9.46  &   9.07  &   4.63  &    12.8  &        \\			
  SCR 0634-5403  &  06 34 36.88 $-$54 03 12.7  &  0.524  &  0.006  &  176.6  &  17.18  &  14.89  &  12.29  &  11.07  &  10.44  &  10.13  &   3.82  &    27.6  &        \\			
  SCR 0642-6707  &  06 42 27.15 $-$67 07 19.9  &  0.811  &  0.008  &  120.4  &  17.00  &  14.69  &  11.60  &  10.61  &  10.15  &   9.81  &   4.08  &    24.1  &        \\			
  SCR 0702-6102  &  07 02 50.33 $-$61 02 47.6  &  0.786  &  0.006  &  041.4  &  17.50  &  15.10  &  11.73  &  10.36  &   9.85  &   9.52  &   4.75  &    15.9  & \tablenotemark{c} \\	      	
  SCR 0723-8015  &  07 23 59.65 $-$80 15 17.8  &  0.828  &  0.006  &  330.4  &  18.68  &  16.44  &  13.27  &  11.30  &  10.82  &  10.44  &   5.14  &    19.3  & \tablenotemark{d} \\	      	
  SCR 0725-8530  &  07 25 22.19 $-$85 30 58.3  &  0.612  &  0.011  &  192.5  &  15.40  &  13.39  &  11.47  &  10.53  &  10.02  &   9.70  &   2.86  &    36.8  &        \\			
  SCR 0730-5707  &  07 30 11.11 $-$57 07 42.4  &  0.505  &  0.007  &  082.7  &  15.08  &  13.03  &  11.23  &  10.23  &   9.73  &   9.47  &   2.80  &    33.8  &        \\			
  SCR 0730-7527  &  07 30 15.98 $-$75 27 29.4  &  0.569  &  0.009  &  000.5  &  14.47  &  12.46  &  11.44  &  11.52  &  11.03  &  10.85  &   0.94  &    99.5  &        \\			
  SCR 0805-5912  &  08 05 46.18 $-$59 12 50.6  &  0.637  &  0.007  &  155.0  &  15.76  &  13.76  &  11.33  &  10.07  &   9.52  &   9.22  &   3.69  &    20.4  &        \\			
  SCR 0816-7727  &  08 16 35.70 $-$77 27 12.0  &  0.676  &  0.006  &  325.4  &  16.41  &  14.43  &  13.58  &  12.62  &  12.07  &  11.87  &   1.81  &   145.7  &        \\			
  SCR 0821-6703  &  08 21 26.67 $-$67 03 20.4  &  0.758  &  0.005  &  327.6  &  16.44  &  15.08  &  14.61  &  13.79  &  13.57  &  13.34  &   1.28  &   267.6  & wd     \\			
  SCR 0829-6203  &  08 29 24.67 $-$62 03 23.2  &  0.585  &  0.005  &  299.2  &  17.78  &  15.72  &  13.42  &  11.69  &  11.21  &  10.92  &   4.02  &    37.9  &        \\			
  SCR 0838-8148  &  08 38 20.47 $-$81 48 46.1  &  0.625  &  0.005  &  009.4  &  17.50  &  15.37  &  13.55  &  12.32  &  11.82  &  11.57  &   3.06  &    77.9  &        \\			
  SCR 0852-6608  &  08 52 49.99 $-$66 08 46.9  &  0.508  &  0.006  &  333.7  &  17.81  &  15.49  &  12.88  &  11.34  &  10.73  &  10.39  &   4.16  &    26.3  &        \\			
  SCR 0853-6123  &  08 53 03.11 $-$61 23 48.4  &  0.587  &  0.006  &  145.7  &  17.93  &  15.73  &  12.96  &  11.82  &  11.27  &  10.91  &   3.91  &    40.3  &        \\			
  SCR 0912-8311  &  09 12 59.55 $-$83 11 51.6  &  0.812  &  0.006  &  331.8  &  17.99  &  15.74  &  13.19  &  11.56  &  10.98  &  10.69  &   4.17  &    30.4  &        \\			
  SCR 0942-6428  &  09 42 17.90 $-$64 28 43.5  &  0.531  &  0.011  &  307.3  &  14.76  &  12.59  &  12.04  &  11.96  &  11.33  &  11.18  &   0.63  &   111.3  &        \\			
  SCR 1057-5103  &  10 57 02.98 $-$51 03 35.0  &  0.622  &  0.009  &  277.2  &  15.96  &  13.85  &  12.20  &  11.15  &  10.64  &  10.43  &   2.70  &    54.1  &        \\			
  SCR 1132-8446  &  11 32 21.98 $-$84 46 28.4  &  0.650  &  0.006  &  279.9  &  17.64  &  15.87  &  13.94  &  12.22  &  11.76  &  11.51  &   3.65  &    62.7  &        \\			
  SCR 1138-7721  &  11 38 16.82 $-$77 21 48.0  &  2.141  &  0.007  &  286.7  &  16.45  &  14.12  &  11.45  &   9.40  &   8.89  &   8.52  &   5.60  &     8.8  & \tablenotemark{e} \\	      	
  SCR 1158-5103  &  11 58 38.90 $-$51 03 31.8  &  0.521  &  0.005  &  294.2  &  18.27  &  16.16  &  14.11  &  13.17  &  12.68  &  12.44  &   2.98  &   123.6  &        \\			
  SCR 1322-7254  &  13 22 27.37 $-$72 54 36.6  &  0.572  &  0.009  &  270.7  &  16.24  &  14.12  &  12.25  &  11.14  &  10.55  &  10.31  &   2.97  &    44.1  &        \\			
  SCR 1328-7253  &  13 28 42.10 $-$72 53 47.4  &  0.789  &  0.005  &  247.2  &  17.91  &  15.97  &  13.77  &  12.47  &  12.00  &  11.69  &   3.50  &    71.2  &        \\			
  SCR 1338-5622  &  13 38 48.13 $-$56 22 20.6  &  0.547  &  0.010  &  260.6  &  15.46  &  14.90  &  13.06  &  13.14  &  12.57  &  12.34  &   1.75  &   170.7  &        \\			
  SCR 1345-5101  &  13 45 41.48 $-$51 01 01.5  &  0.527  &  0.006  &  168.4  &  16.61  &  14.51  &  12.17  &  10.91  &  10.39  &  10.12  &   3.60  &    31.9  &        \\			
  SCR 1429-4808  &  14 29 41.38 $-$48 08 31.2  &  0.791  &  0.006  &  351.4  &  17.11  &  15.48  &  12.52  &  11.25  &  10.78  &  10.45  &   4.24  &    32.8  &        \\			
  SCR 1442-4810  &  14 42 16.59 $-$48 10 50.8  &  0.507  &  0.008  &  248.0  &  16.92  &  14.33  &  14.13  &  12.98  &  12.47  &  12.29  &   1.35  &   173.8  &        \\			
  SCR 1659-6958  &  16 59 27.99 $-$69 58 18.7  &  0.749  &  0.007  &  216.3  &  15.77  &  14.19  &  12.02  &  10.53  &   9.99  &   9.70  &   3.65  &    27.9  &        \\			
  SCR 1726-8433  &  17 26 23.04 $-$84 33 08.4  &  0.518  &  0.008  &  134.8  &  15.42  &  13.31  &  11.16  &   9.87  &   9.33  &   9.02  &   3.44  &    20.1  &        \\			
  SCR 1735-7020  &  17 35 40.71 $-$70 20 21.6  &  0.963  &  0.005  &  190.1  &  18.19  &  16.14  &  14.04  &  12.82  &  12.31  &  12.10  &   3.32  &    90.9  &        \\			
  SCR 1756-5927  &  17 56 27.94 $-$59 27 18.0  &  0.537  &  0.006  &  210.0  &  18.02  &  15.73  &  14.68  &  13.44  &  12.89  &  12.69  &   2.29  &   170.6  &        \\			
  SCR 1808-8120  &  18 08 00.06 $-$81 20 48.8  &  0.680  &  0.009  &  200.0  &  17.32  &  15.45  &  13.19  &  11.36  &  10.79  &  10.52  &   4.09  &    31.0  &        \\			
  SCR 1817-5318  &  18 17 06.43 $-$53 18 04.8  &  0.617  &  0.009  &  209.9  &  14.69  &  13.27  &  12.46  &  11.93  &  11.43  &  11.23  &   1.34  &   114.6  &        \\			
  SCR 1835-8754  &  18 35 14.60 $-$87 54 08.9  &  0.639  &  0.006  &  199.5  &  18.18  &  16.02  &  15.10  &  14.11  &  13.56  &  13.29  &   1.92  &   264.2  &        \\			
  SCR 1843-7849  &  18 43 35.71 $-$78 49 02.6  &  0.745  &  0.008  &  194.9  &  17.57  &  15.70  &  14.65  &  13.27  &  12.74  &  12.59  &   2.43  &   168.4  &        \\			
  SCR 1845-6357  &  18 45 05.09 $-$63 57 47.7  &  2.558  &  0.012  &  074.8  &  \nodata&  16.33  &  12.53  &   9.54  &   8.97  &   8.51  &   6.78  &     3.5  & \tablenotemark{f} \\	      	
  SCR 1848-6855  &  18 48 21.14 $-$68 55 34.5  &  1.287  &  0.013  &  194.3  &  \nodata&  16.07  &  13.97  &  11.89  &  11.40  &  11.10  &   4.97  &    34.8  & \tablenotemark{g} \\	      	
  SCR 1855-6914  &  18 55 47.87 $-$69 14 14.8  &  0.832  &  0.011  &  145.3  &  18.01  &  15.63  &  12.20  &  10.47  &   9.88  &   9.51  &   5.16  &    12.5  &        \\			
  SCR 1902-5044  &  19 02 47.53 $-$50 44 00.6  &  0.510  &  0.009  &  150.2  &  16.54  &  14.52  &  12.98  &  11.99  &  11.48  &  11.26  &   2.53  &    86.7  &        \\			
  SCR 1946-4945  &  19 46 02.47 $-$49 45 49.0  &  0.585  &  0.006  &  210.2  &  17.34  &  15.39  &  14.53  &  13.51  &  12.95  &  12.78  &   1.88  &   218.5  &        \\			
  SCR 1954-7356  &  19 54 06.43 $-$73 56 50.8  &  0.535  &  0.008  &  148.6  &  16.85  &  14.89  &  12.96  &  11.81  &  11.31  &  11.08  &   3.08  &    64.3  &        \\			
  SCR 2012-5956  &  20 12 31.79 $-$59 56 51.6  &  1.440  &  0.011  &  165.6  &  16.66  &  15.63  &  15.13  &  14.93  &  15.23  &  15.41  &   0.70  &   734.1  & wd, \tablenotemark{h} \\	      	
  SCR 2035-6505  &  20 35 05.60 $-$65 05 26.1  &  0.785  &  0.011  &  166.0  &  17.22  &  15.07  &  13.18  &  12.23  &  11.73  &  11.51  &   2.85  &    84.3  &        \\			
  SCR 2040-5501  &  20 40 12.40 $-$55 01 25.7  &  0.514  &  0.013  &  125.4  &  16.56  &  14.26  &  12.16  &  10.56  &  10.02  &   9.69  &   3.70  &    22.9  &        \\			
  SCR 2043-6501  &  20 43 10.43 $-$65 01 17.6  &  0.533  &  0.013  &  170.0  &  16.18  &  14.04  &  12.04  &  11.25  &  10.76  &  10.52  &   2.79  &    55.3  &        \\			
  SCR 2101-5437  &  21 01 45.76 $-$54 37 31.7  &  0.667  &  0.011  &  241.5  &  16.90  &  14.59  &  13.46  &  12.79  &  12.26  &  12.08  &   1.80  &   157.0  &        \\			
  SCR 2109-5226  &  21 09 02.56 $-$52 26 18.1  &  0.791  &  0.012  &  176.5  &  18.00  &  15.97  &  14.93  &  13.76  &  13.29  &  13.05  &   2.21  &   221.8  &        \\			
  SCR 2128-5532  &  21 28 41.23 $-$55 32 32.1  &  0.699  &  0.010  &  123.3  &  16.37  &  14.23  &  12.04  &  10.70  &  10.06  &   9.78  &   3.53  &    26.4  &        \\			
  SCR 2130-7710  &  21 30 07.07 $-$77 10 37.5  &  0.589  &  0.007  &  118.0  &  18.28  &  15.93  &  13.44  &  11.29  &  10.67  &  10.36  &   4.64  &    20.6  &        \\			
  SCR 2235-7722  &  22 35 57.78 $-$77 22 16.2  &  0.612  &  0.009  &  197.6  &  18.42  &  16.36  &  \nodata&  14.17  &  13.67  &  13.46  &   2.19  &   285.9  &        \\			
  SCR 2250-5726AB&  22 50 45.05 $-$57 26 01.8  &  0.714  &  0.007  &  117.3  &  18.07  &  16.10  &  13.80  &  12.63  &  12.00  &  11.81  &   3.48  &    73.7  & \tablenotemark{i} \\	      	
  SCR 2307-8452  &  23 07 19.88 $-$84 52 03.8  &  0.613  &  0.011  &  097.2  &  16.33  &  14.16  &  11.83  &  10.36  &   9.81  &   9.47  &   3.80  &    20.6  &        \\			
  SCR 2335-5020  &  23 35 52.96 $-$50 20 18.9  &  0.661  &  0.010  &  127.0  &  16.54  &  15.17  &  13.97  &  13.14  &  12.69  &  12.47  &   2.03  &   179.4  &        \\			
  SCR 2352-6124  &  23 52 29.56 $-$61 24 23.1  &  0.848  &  0.009  &  167.1  &  17.10  &  14.73  &  12.63  &  11.52  &  11.02  &  10.82  &   3.21  &    50.3  & cpm with LHS 4031  \\	      	
\tableline \vspace{-15pt} \\															     					
\multicolumn{14}{c}{SuperCOSMOS$-$RECONS sample with $\mu$ between 0.4$\arcsec$/yr and 0.5$\arcsec$/yr} \\					     					
\tableline \vspace{-15pt} \\															     					
  SCR 0000-5029  &  00 00 44.12 $-$50 29 25.0  &  0.402  &  0.017  &  091.8  &  15.55  &  13.44  &  11.66  &  11.22  &  10.73  &  10.49  &   2.22  &    68.3  &        \\			
  SCR 0122-6400  &  01 22 21.37 $-$64 00 33.1  &  0.423  &  0.012  &  113.9  &  15.13  &  13.23  &  12.48  &  12.53  &  11.93  &  11.80  &   0.70  &   176.0  &        \\			
  SCR 0128-7104  &  01 28 50.80 $-$71 04 52.7  &  0.452  &  0.010  &  088.8  &  17.59  &  15.45  &  13.23  &  12.64  &  12.13  &  11.88  &   2.81  &   103.5  &        \\			
  SCR 0133-7200  &  01 33 13.09 $-$72 00 04.6  &  0.433  &  0.008  &  172.2  &  16.92  &  14.64  &  12.16  &  11.37  &  10.79  &  10.50  &   3.27  &    43.0  &        \\			
  SCR 0135-5943  &  01 35 46.71 $-$59 43 14.3  &  0.412  &  0.008  &  081.5  &  17.36  &  15.25  &  13.01  &  12.01  &  11.52  &  11.24  &   3.24  &    63.7  &        \\			
  SCR 0149-8038  &  01 49 43.55 $-$80 38 27.8  &  0.464  &  0.008  &  080.5  &  18.42  &  16.35  &  13.84  &  11.68  &  11.11  &  10.72  &   4.68  &    25.3  &        \\			
  SCR 0210-6252  &  02 10 43.99 $-$62 52 30.1  &  0.456  &  0.018  &  050.0  &  17.23  &  14.95  &  12.92  &  11.85  &  11.29  &  11.02  &   3.10  &    56.8  &        \\			
  SCR 0224-6433  &  02 24 10.98 $-$64 33 02.4  &  0.448  &  0.023  &  107.3  &  16.26  &  13.97  &  11.80  &  10.94  &  10.43  &  10.12  &   3.03  &    39.9  &        \\			
  SCR 0242-5935  &  02 42 26.34 $-$59 35 02.4  &  0.466  &  0.007  &  185.2  &  16.04  &  15.02  &  14.04  &  13.55  &  13.00  &  12.78  &   1.46  &   228.9  &        \\			
  SCR 0252-7522  &  02 52 45.57 $-$75 22 44.5  &  0.496  &  0.013  &  063.5  &  17.10  &  16.32  &  16.17  &  15.77  &  15.76  &  15.34  &   0.55  &   675.9  & wd     \\			
  SCR 0255-7242  &  02 55 05.52 $-$72 42 42.1  &  0.439  &  0.013  &  051.7  &  17.52  &  15.44  &  14.26  &  13.74  &  13.23  &  13.01  &   1.70  &   254.4  &        \\			
  SCR 0303-7209  &  03 03 44.13 $-$72 09 59.9  &  0.430  &  0.009  &  085.9  &  18.77  &  16.38  &  14.11  &  12.72  &  12.23  &  11.95  &   3.66  &    67.8  &        \\			
  SCR 0311-6215  &  03 11 21.28 $-$62 15 15.9  &  0.416  &  0.015  &  083.3  &  15.68  &  16.05  &  16.13  &  16.13  &  16.31  &  16.50  &  -0.08  &  \nodata & wd, \tablenotemark{j} \\	      	
  SCR 0331-8251  &  03 31 41.78 $-$82 51 10.5  &  0.447  &  0.007  &  050.8  &  18.24  &  16.43  &  14.66  &  13.21  &  12.69  &  12.46  &   3.22  &   115.2  &        \\			
  SCR 0525-7425  &  05 25 45.56 $-$74 25 25.9  &  0.417  &  0.009  &  040.2  &  14.81  &  12.89  &  11.35  &  10.03  &   9.42  &   9.21  &   2.86  &    28.7  &        \\			
  SCR 0537-5612  &  05 37 53.75 $-$56 12 17.4  &  0.402  &  0.009  &  122.7  &  \nodata&  14.89  &  13.12  &  12.34  &  11.85  &  11.57  &   2.56  &    96.9  &        \\			
  SCR 0615-5807  &  06 15 05.02 $-$58 07 43.4  &  0.410  &  0.006  &  314.6  &  16.63  &  14.45  &  12.44  &  11.48  &  10.98  &  10.70  &   2.97  &    54.4  &        \\			
  SCR 0618-6704  &  06 18 26.01 $-$67 04 00.3  &  0.436  &  0.009  &  031.4  &  14.59  &  12.67  &  10.74  &  10.40  &   9.88  &   9.60  &   2.27  &    45.7  &        \\			
  SCR 0629-6938  &  06 29 56.40 $-$69 38 13.3  &  0.473  &  0.007  &  153.6  &  18.14  &  16.23  &  14.75  &  13.66  &  13.14  &  12.90  &   2.57  &   183.8  &        \\			
  SCR 0630-7643AB&  06 30 46.63 $-$76 43 09.2  &  0.483  &  0.008  &  356.8  &  15.78  &  13.56  &  10.74  &   8.89  &   8.27  &   7.92  &   4.67  &     6.9  & \tablenotemark{k} \\	      	
  SCR 0654-7358  &  06 54 06.34 $-$73 58 04.0  &  0.467  &  0.008  &  020.2  &  18.19  &  16.24  &  15.11  &  13.99  &  13.47  &  13.28  &   2.25  &   246.7  &        \\			
  SCR 0740-7212  &  07 40 00.80 $-$72 12 27.8  &  0.481  &  0.006  &  003.7  &  17.25  &  15.28  &  13.31  &  11.77  &  11.27  &  11.00  &   3.51  &    49.8  &        \\			
  SCR 0744-6941  &  07 44 35.21 $-$69 41 58.1  &  0.441  &  0.007  &  001.2  &  17.14  &  15.05  &  13.38  &  12.18  &  11.69  &  11.41  &   2.87  &    78.8  &        \\			
  SCR 0756-5434  &  07 56 48.71 $-$54 34 57.1  &  0.446  &  0.005  &  324.2  &  17.86  &  15.91  &  13.56  &  11.86  &  11.28  &  10.98  &   4.05  &    38.2  &        \\			
  SCR 0812-6402  &  08 12 23.36 $-$64 02 24.0  &  0.409  &  0.005  &  340.0  &  17.08  &  15.23  &  13.26  &  11.80  &  11.32  &  11.03  &   3.43  &    54.4  &        \\			
  SCR 0824-6721  &  08 24 03.20 $-$67 21 50.5  &  0.403  &  0.005  &  288.9  &  17.54  &  15.29  &  13.13  &  11.55  &  10.95  &  10.70  &   3.74  &    36.0  &        \\			
  SCR 0843-5154  &  08 43 11.02 $-$51 54 03.4  &  0.402  &  0.008  &  310.7  &  16.10  &  13.94  &  12.36  &  11.77  &  11.22  &  10.97  &   2.18  &    84.6  &        \\			
  SCR 0843-5209  &  08 43 38.80 $-$52 09 27.5  &  0.482  &  0.007  &  307.3  &  16.41  &  14.29  &  12.57  &  11.87  &  11.38  &  11.11  &   2.42  &    84.2  &        \\			
  SCR 0850-4934  &  08 50 24.90 $-$49 34 23.7  &  0.469  &  0.006  &  295.5  &  17.42  &  15.60  &  13.46  &  12.22  &  11.77  &  11.52  &   3.38  &    72.3  &        \\			
  SCR 0956-8518  &  09 56 14.12 $-$85 18 01.5  &  0.478  &  0.007  &  319.2  &  17.11  &  15.14  &  13.98  &  12.49  &  11.94  &  11.74  &   2.64  &   100.7  &        \\			
  SCR 1011-8106  &  10 11 12.37 $-$81 06 42.0  &  0.450  &  0.008  &  112.4  &  16.54  &  14.50  &  12.53  &  10.81  &  10.24  &   9.93  &   3.68  &    26.7  &        \\			
  SCR 1054-5159  &  10 54 16.31 $-$51 59 03.0  &  0.408  &  0.006  &  306.1  &  17.32  &  15.38  &  13.37  &  11.71  &  11.10  &  10.87  &   3.66  &    42.7  &        \\			
  SCR 1104-8352  &  11 04 51.06 $-$83 52 25.2  &  0.440  &  0.015  &  256.7  &  \nodata&  13.47  &  12.23  &  10.53  &   9.96  &   9.67  &   2.94  &    29.1  &        \\			
  SCR 1143-7047  &  11 43 11.44 $-$70 47 21.4  &  0.460  &  0.007  &  266.5  &  17.58  &  15.63  &  13.81  &  12.35  &  11.82  &  11.59  &   3.28  &    72.5  &        \\			
  SCR 1155-7904  &  11 55 00.07 $-$79 04 13.1  &  0.401  &  0.006  &  297.3  &  18.17  &  16.23  &  14.99  &  13.30  &  12.66  &  12.44  &   2.94  &   119.6  &        \\			
  SCR 1211-6849  &  12 11 39.70 $-$68 49 29.9  &  0.489  &  0.008  &  293.4  &  16.91  &  14.73  &  12.43  &  11.39  &  10.91  &  10.62  &   3.35  &    45.4  &        \\			
  SCR 1213-4820  &  12 13 07.11 $-$48 20 07.9  &  0.480  &  0.006  &  268.0  &  16.51  &  14.24  &  11.96  &  11.25  &  10.72  &  10.45  &   2.99  &    47.9  &        \\			
  SCR 1239-4759  &  12 39 51.37 $-$47 59 07.8  &  0.401  &  0.006  &  268.5  &  17.26  &  15.26  &  12.92  &  11.57  &  11.08  &  10.80  &   3.69  &    42.8  &        \\			
  SCR 1240-8116  &  12 40 56.05 $-$81 16 31.1  &  0.492  &  0.006  &  279.8  &  15.15  &  13.12  &  11.25  &   9.73  &   9.16  &   8.89  &   3.39  &    19.2  &        \\			
  SCR 1240-8209  &  12 40 51.09 $-$82 09 03.4  &  0.486  &  0.008  &  272.4  &  16.18  &  14.56  &  12.30  &  10.85  &  10.20  &   9.93  &   3.70  &    29.3  &        \\			
  SCR 1245-5506  &  12 45 52.60 $-$55 06 49.9  &  0.412  &  0.011  &  107.0  &  14.84  &  12.82  &  10.34  &   8.99  &   8.43  &   8.12  &   3.83  &    11.5  &        \\			
  SCR 1257-5554A &  12 57 32.84 $-$55 54 48.6  &  0.410  &  0.012  &  290.1  &  14.84  &  13.47  &  11.45  &  10.48  &   9.90  &   9.66  &   2.98  &    39.1  & \tablenotemark{l} \\	      	
  SCR 1257-5554B &  12 57 33.08 $-$55 54 38.0  &  0.403  &  0.006  &  293.2  &  17.30  &  16.94  &  16.82  &  \nodata&  \nodata&  \nodata&  \nodata&  \nodata & \tablenotemark{m} \\	      	
  SCR 1320-7542  &  13 20 47.55 $-$75 42 51.1  &  0.434  &  0.006  &  249.3  &  17.88  &  15.82  &  14.71  &  13.93  &  13.32  &  13.26  &   1.89  &   270.9  &        \\			
  SCR 1331-5138  &  13 31 06.82 $-$51 38 02.8  &  0.484  &  0.007  &  294.9  &  16.07  &  14.10  &  11.95  &  10.99  &  10.50  &  10.27  &   3.11  &    44.5  &        \\			
  SCR 1409-5337  &  14 09 49.48 $-$53 37 26.4  &  0.450  &  0.007  &  212.4  &  16.45  &  14.33  &  13.06  &  11.72  &  11.24  &  10.96  &   2.61  &    70.2  &        \\			
  SCR 1412-4954  &  14 12 43.89 $-$49 54 32.4  &  0.420  &  0.012  &  212.9  &  15.86  &  13.82  &  12.25  &  11.66  &  11.12  &  10.89  &   2.16  &    83.9  &        \\			
  SCR 1420-5106  &  14 20 21.71 $-$51 06 50.7  &  0.489  &  0.010  &  130.4  &  14.64  &  13.02  &  11.57  &  10.47  &   9.92  &   9.70  &   2.55  &    44.5  &        \\			
  SCR 1443-5502  &  14 43 25.99 $-$55 02 53.0  &  0.477  &  0.017  &  277.5  &  15.08  &  13.22  &  11.57  &  10.28  &   9.71  &   9.49  &   2.94  &    32.4  &        \\			
  SCR 1445-5046  &  14 45 23.96 $-$50 46 06.4  &  0.435  &  0.007  &  244.1  &  17.10  &  15.52  &  13.64  &  12.02  &  11.50  &  11.30  &   3.49  &    63.1  &        \\			
  SCR 1552-7052  &  15 52 46.95 $-$70 52 02.4  &  0.468  &  0.006  &  216.0  &  16.33  &  14.18  &  11.92  &  10.81  &  10.28  &  10.07  &   3.37  &    34.5  &        \\			
  SCR 1627-7337  &  16 27 37.13 $-$73 37 06.2  &  0.439  &  0.007  &  235.8  &  15.71  &  13.89  &  13.26  &  12.65  &  12.02  &  11.95  &   1.23  &   151.7  &        \\			
  SCR 1717-6916  &  17 17 52.66 $-$69 16 43.2  &  0.466  &  0.005  &  320.9  &  17.45  &  15.04  &  12.86  &  11.48  &  10.94  &  10.67  &   3.56  &    38.4  &        \\			
  SCR 1724-5637  &  17 24 36.47 $-$56 37 02.9  &  0.421  &  0.008  &  154.5  &  15.18  &  13.11  &  11.42  &  10.39  &   9.86  &   9.63  &   2.72  &    37.2  &        \\			
  SCR 1739-8222  &  17 39 45.45 $-$82 22 02.3  &  0.465  &  0.008  &  211.8  &  17.06  &  15.04  &  14.21  &  12.90  &  12.38  &  12.19  &   2.14  &   151.7  &        \\			
  SCR 1740-5646  &  17 40 46.93 $-$56 46 57.9  &  0.448  &  0.005  &  229.5  &  18.48  &  15.90  &  14.44  &  13.83  &  13.33  &  13.19  &   2.07  &   232.9  &        \\			
  SCR 1748-7211  &  17 48 51.85 $-$72 11 53.1  &  0.428  &  0.008  &  191.0  &  16.99  &  14.85  &  12.88  &  11.57  &  10.99  &  10.75  &   3.28  &    46.9  &        \\			
  SCR 1811-5510  &  18 11 34.94 $-$55 10 37.9  &  0.482  &  0.006  &  197.9  &  17.39  &  15.16  &  12.87  &  11.62  &  11.06  &  10.78  &   3.54  &    42.6  &        \\			
  SCR 1821-5549  &  18 21 45.87 $-$55 49 17.5  &  0.424  &  0.007  &  181.0  &  16.97  &  14.63  &  12.61  &  11.57  &  11.00  &  10.74  &   3.06  &    50.2  &        \\			
  SCR 1912-5034  &  19 12 45.01 $-$50 34 34.4  &  0.447  &  0.005  &  233.7  &  18.24  &  16.03  &  13.70  &  12.19  &  11.67  &  11.35  &   3.84  &    48.3  &        \\			
  SCR 1926-5218  &  19 26 48.73 $-$52 18 17.4  &  0.494  &  0.007  &  191.2  &  16.97  &  15.22  &  14.39  &  13.54  &  12.97  &  12.87  &   1.68  &   229.7  &        \\			
  SCR 1931-5840  &  19 31 21.58 $-$58 40 37.2  &  0.402  &  0.007  &  135.7  &  16.83  &  14.69  &  13.32  &  12.18  &  11.67  &  11.44  &   2.51  &    91.7  &        \\			
  SCR 1948-5914  &  19 48 58.82 $-$59 14 23.3  &  0.415  &  0.008  &  151.7  &  16.60  &  14.63  &  12.39  &  11.11  &  10.58  &  10.28  &   3.52  &    35.9  &        \\			
  SCR 1958-5609  &  19 58 31.28 $-$56 09 10.6  &  0.494  &  0.007  &  161.9  &  17.60  &  15.55  &  14.41  &  13.30  &  12.77  &  12.52  &   2.25  &   169.1  &        \\			
  SCR 1959-5549  &  19 59 58.76 $-$55 49 29.6  &  0.413  &  0.011  &  169.9  &  16.19  &  13.95  &  11.82  &  10.47  &   9.88  &   9.63  &   3.48  &    25.0  &        \\			
  SCR 2009-6005  &  20 09 23.44 $-$60 05 43.3  &  0.414  &  0.010  &  154.3  &  18.17  &  15.87  &  13.71  &  12.11  &  11.58  &  11.27  &   3.76  &    46.5  &        \\			
  SCR 2016-7945  &  20 16 49.73 $-$79 45 53.0  &  0.434  &  0.007  &  128.4  &  16.75  &  16.09  &  15.75  &  15.11  &  15.03  &  14.64  &   0.99  &   482.5  & wd     \\			
  SCR 2018-4836  &  20 18 13.66 $-$48 36 51.9  &  0.410  &  0.007  &  147.2  &  17.38  &  15.20  &  13.33  &  12.08  &  11.62  &  11.37  &   3.12  &    69.0  &        \\			
  SCR 2018-6606  &  20 18 28.69 $-$66 06 44.5  &  0.462  &  0.008  &  191.3  &  17.72  &  15.76  &  14.71  &  13.68  &  13.14  &  12.99  &   2.08  &   228.0  &        \\			
  SCR 2104-5229  &  21 04 00.61 $-$52 29 43.4  &  0.400  &  0.009  &  233.7  &  17.60  &  15.42  &  14.42  &  13.44  &  12.94  &  12.76  &   1.98  &   207.2  &        \\			
  SCR 2151-8604  &  21 51 37.56 $-$86 04 33.4  &  0.454  &  0.012  &  192.7  &  16.57  &  14.48  &  13.57  &  12.74  &  12.23  &  12.03  &   1.74  &   160.6  &        \\			
  SCR 2155-7330  &  21 55 47.55 $-$73 30 24.5  &  0.459  &  0.011  &  202.0  &  15.78  &  13.97  &  11.85  &  10.60  &  10.05  &   9.78  &   3.38  &    31.6  & cpm with HIP 108158   \\	      	
  SCR 2212-7337  &  22 12 05.47 $-$73 37 17.2  &  0.419  &  0.008  &  122.9  &  17.43  &  15.32  &  12.90  &  11.39  &  10.85  &  10.48  &   3.92  &    31.6  &        \\			
  SCR 2249-6324  &  22 49 47.16 $-$63 24 37.7  &  0.454  &  0.009  &  174.0  &  18.26  &  16.28  &  15.50  &  14.69  &  14.04  &  13.95  &   1.58  &   393.6  &        \\			
  SCR 2254-8712  &  22 54 21.45 $-$87 12 51.7  &  0.401  &  0.011  &  115.6  &  16.11  &  14.13  &  12.30  &  11.10  &  10.54  &  10.29  &   3.03  &    44.2  &        \\			
  SCR 2305-7729  &  23 05 01.97 $-$77 29 12.8  &  0.429  &  0.007  &  193.7  &  17.31  &  15.73  &  14.89  &  13.82  &  13.30  &  13.18  &   1.91  &   238.7  &        \\			
  SCR 2317-5140  &  23 17 08.89 $-$51 40 19.4  &  0.446  &  0.007  &  192.1  &  17.04  &  15.02  &  13.71  &  12.82  &  12.25  &  12.04  &   2.19  &   139.4  &        \\			
  SCR 2329-8758  &  23 29 02.47 $-$87 58 06.2  &  0.429  &  0.006  &  111.6  &  15.79  &  14.48  &  13.44  &  12.70  &  12.12  &  11.97  &   1.77  &   144.5  &        \\                        

\enddata			     

\tablenotetext{a}{first reported in Hambly et al.~2004; 38.1 $\pm$ 7.8 pc in Henry et al.~2004}
\tablenotetext{b}{15.4 $\pm$ 2.6 pc in Henry et al.~2004}
\tablenotetext{c}{10.8 $\pm$ 2.1 pc in Henry et al.~2004}
\tablenotetext{d}{17.2 $\pm$ 3.1 pc in Henry et al.~2004}
\tablenotetext{e}{first reported in Hambly et al.~2004; 9.4 $\pm$ 1.7 pc in Henry et al.~2004}
\tablenotetext{f}{first reported in Hambly et al.~2004; 4.6 $\pm$ 0.8 pc in Henry et al.~2004}
\tablenotetext{g}{first reported in Hambly et al.~2004; 37.0 $\pm$ 9.4 pc in Henry et al.~2004}
\tablenotetext{h}{first reported in Hambly et al.~2004; 17.4 $\pm$ 3.5 pc in Henry et al.~2004; K$_s$ suspect}
\tablenotetext{i}{separation 2.3$\arcsec$ at PA 28$^\circ$}
\tablenotetext{j}{all colors too blue for distance relations}
\tablenotetext{k}{7.0 $\pm$ 1.2 pc in Henry et al.~2004, binary with separation $\sim$1.0$\arcsec$}
\tablenotetext{l}{separation 11.0$\arcsec$ at PA 16$^\circ$}
\tablenotetext{m}{not detected during automated search due to faint limit but noticed to be a common proper motion companion during the visual inspection; R is $ESO-R$, $R_{59F}$ blended}

\end{deluxetable}


\begin{thebibliography}{}

\bibitem[Deacon et al.(2005)]{dea05} Deacon, N.~R., Hambly, N.~C.,
Henry, T.~J., Subasavage, J.~P., Brown, M.~A., \& Jao, W.~C.\ 2005,
\aj, submitted

\bibitem[Esa(1997)]{1997yCat.1239....0E} Esa, 1.\ 1997, VizieR Online
Data Catalog, 1239, 0

\bibitem[Giclas et al.(1971)]{1971lpms.book.....G}
Giclas, H.~L., Burnham, R., \& Thomas, N.~G.\ 1971, Flagstaff,
Arizona: Lowell Observatory, 1971

\bibitem[Giclas et al.(1978)]{1978LowOB...8...89G}
Giclas, H.~L., Burnham, R., \& Thomas, N.~G.\ 1978, Lowell Observatory
Bulletin, 8, 89

\bibitem[Gliese \& Jahrei{\ss}(1991)]{1991adc..rept.....G} Gliese,
W.~\& Jahrei{\ss}, H.\ 1991, On: The Astronomical Data Center CD-ROM:
Selected Astronomical Catalogs, Vol.~I; L.E.~Brotzmann, S.E.~Gesser
(eds.), NASA/Astronomical Data Center, Goddard Space Flight Center,
Greenbelt, MD

\bibitem[Hambly et al.(2001)]{2001MNRAS.326.1295H} Hambly, N.~C.,
Irwin, M.~J., \& MacGillivray, H.~T.\ 2001, \mnras, 326, 1295

\bibitem[Hambly et al.(2004)]{2004AJ....128..437H} Hambly, N.~C.,
Henry, T.~J., Subasavage, J.~P., Brown, M.~A., \& Jao, W.\ 2004, \aj,
128, 437

\bibitem[Henry et al.(1997)]{1997AJ....114..388H} Henry, T.~J., Ianna, P.~A.,
Kirkpatrick, J.~D., \& Jahreiss, H.\ 1997, \aj, 114, 388

\bibitem[Henry et al.(2004)] {hen04} Henry, T.~J., Subasavage, J.~P.,
Brown, M.~A., Beaulieu, T.~D., Jao, W.~C., \& Hambly, N.~C.\ 2004,
\aj, submitted

\bibitem[Holberg et al.(2002)]{2002ApJ...571..512H} Holberg,
J.~B., Oswalt, T.~D., \& Sion, E.~M.\ 2002, \apj, 571, 512

\bibitem[L{\' e}pine et al.(2002)]{2002AJ....124.1190L} L{\' 
e}pine, S., Shara, M.~M., \& Rich, R.~M.\ 2002, \aj, 124, 1190 

\bibitem[L{\' e}pine et al.(2003)]{2003AJ....126..921L} L{\'
e}pine, S., Shara, M.~M., \& Rich, R.~M.\ 2003, \aj, 126, 921

\bibitem[Luyten(1979)]{1979lccs.book.....L} Luyten, W.~J.\ 1979, 
Minneapolis: University of Minnesota, 1979, 2nd ed., 

\bibitem[Luyten(1995)]{1995yCat.1098....0L} Luyten, W.~J.\ 1995,
VizieR Online Data Catalog, 1098, 0

\bibitem[Oppenheimer et al.(2001)]{2001Sci...292..698O} Oppenheimer,
B.~R., Hambly, N.~C., Digby, A.~P., Hodgkin, S.~T., \& Saumon, D.\
2001, Science, 292, 698

\bibitem[Pokorny et al.(2003)]{2003A&A...397..575P} Pokorny, R.~S.,
Jones, H.~R.~A., \& Hambly, N.~C.\ 2003, \aap, 397, 575

\bibitem[Ruiz \& Maza(1987)]{1987RMxAA..14..381R} Ruiz, M.~T.~\& Maza,
J.\ 1987, Revista Mexicana de Astronomia y Astrofisica, vol.~14, 14,
381

\bibitem[Ruiz et al.(2001)]{2001ApJS..133..119R} Ruiz, M.~T.,
Wischnjewsky, M., Rojo, P.~M., \& Gonzalez, L.~E.\ 2001, \apjs, 133,
119

\bibitem[Salim et al.(2004)]{2004ApJ...601.1075S} Salim, S., Rich, R.~M., 
Hansen, B.~M., Koopmans, L.~V.~E., Oppenheimer, B.~R., \& Blandford, R.~D.\ 
2004, \apj, 601, 1075

\bibitem[Scholz et al.(2000)]{2000A&A...353..958S} Scholz, R.-D., Irwin, 
M., Ibata, R., Jahrei{\ss}, H., \& Malkov, O.~Y.\ 2000, \aap, 353, 958

\bibitem[Scholz et al.(2002)]{2002ApJ...565..539S} Scholz, R.-D., Szokoly, 
G.~P., Andersen, M., Ibata, R., \& Irwin, M.~J.\ 2002, \apj, 565, 539

\bibitem[Scholz et al.(2004)]{astro-ph/0406457} Scholz, R.-D.,
Lehmann, I., Matute, I., \& Zinnecker, H.\ 2004, \aap, preprint
(astro-ph/0406457)

\bibitem[van Altena et al.(1995)]{1995gcts.book.....V} van
Altena, W.~F., Lee, J.~T., \& Hoffleit, E.~D.\ 1995, New Haven, CT:
Yale University Observatory, [1995, 4th ed., completely revised and
enlarged]

\bibitem[Wroblewski \& Torres(1989)]{1989A&AS...78..231W} Wroblewski,
H.~\& Torres, C.\ 1989, \aaps, 78, 231

\bibitem[Wroblewski \& Torres(1991)]{1991A&AS...91..129W} Wroblewski,
H.~\& Torres, C.\ 1991, \aaps, 91, 129

\bibitem[Wroblewski \& Torres(1994)]{1994A&AS..105..179W} Wroblewski,
H.~\& Torres, C.\ 1994, \aaps, 105, 179

\bibitem[Wroblewski \& Torres(1996)]{1996A&AS..115..481W} Wroblewski,
H.~\& Torres, C.\ 1996, \aaps, 115, 481

\bibitem[Wroblewski \& Torres(1997)]{1997A&AS..122..447W} Wroblewski,
H.~\& Torres, C.\ 1997, \aaps, 122, 447

\bibitem[Wroblewski \& Costa(1999)]{1999A&AS..139...25W} Wroblewski,
H.~\& Costa, E.\ 1999, \aaps, 139, 25

\bibitem[Wroblewski \& Costa(2001)]{2001A&A...367..725W} Wroblewski,
H.~\& Costa, E.\ 2001, \aap, 367, 725

\end{thebibliography}
\end{document}